\documentclass[review,3p]{elsarticle}

\usepackage{lineno}
\usepackage{amsmath}
\modulolinenumbers[5]
\usepackage{todonotes}
\presetkeys
    {todonotes}
    {inline,backgroundcolor=yellow}{}
\usepackage{algorithm}
\usepackage{algorithmicx}
\usepackage{algpseudocode}
\usepackage{mathtools}

\usepackage{amssymb}
\usepackage{csquotes}
\usepackage{graphicx}
\usepackage{booktabs}
\usepackage[normalem]{ulem}
\usepackage[flushleft]{threeparttable}
\usepackage{appendix}
\usepackage{verbatim}
\usepackage{subcaption}

 \usepackage{hyperref}
 \hypersetup{
    colorlinks=false,
    linkcolor=blue,
    filecolor=magenta,
    urlcolor=blue,
    citecolor=blue
    }

\modulolinenumbers[5]

\journal{arXiv}

\biboptions{authoryear}

\begin{document}

\begin{frontmatter}

\title{Balancing Consumer and Business Value of Recommender Systems: A Simulation-based Analysis}

\author{Nada Ghanem}
\ead{nada.ghanem@aau.at}

\author{Stephan Leitner\corref{mycorrespondingauthor}}
\cortext[mycorrespondingauthor]{Corresponding author. Tel.: +43 (0) 463 2700 4035; fax: +43 (0) 463 2700 994035}
\ead{stephan.leitner@aau.at}

\author{Dietmar Jannach}
\ead{dietmar.jannach@aau.at}

\address{University of Klagenfurt, Universit\"atsstra{\ss}e 65-67, 9020 Klagenfurt, Austria}

\begin{abstract}
Automated recommendations can nowadays be found on many e-commerce platforms,
and such recommendations can create substantial value for consumers and providers. Often, however, not all recommendable items have the same profit margin, and providers might thus be tempted to promote items that maximize their profit. In the short run, consumers might accept non-optimal recommendations, but they may lose their trust in the long run.
Ultimately, this leads to the problem of designing  balanced recommendation strategies, which consider both consumer and provider value and lead to sustained business success.

This work proposes a simulation framework based on agent-based modeling designed to help providers explore longitudinal dynamics of different recommendation strategies. In our model, consumer agents receive recommendations from providers, and the perceived quality of the recommendations influences the consumers' trust over time.
\textcolor{black}{We design several recommendation strategies which either give more weight on provider profit or on consumer utility. Our simulations show that a hybrid strategy that puts more weight on consumer utility but without ignoring profitability considerations leads to the highest cumulative profit in the long run. This hybrid strategy results in a profit increase of about 20\,\% compared to pure consumer or profit oriented strategies. We also find that social media can reinforce the observed phenomena. In case when consumers heavily rely on social media, the cumulative profit of the best strategy further increases.} To ensure reproducibility and foster future research, we publicly share our flexible simulation framework.

\end{abstract}


\end{frontmatter}

\section{Introduction}
Recommender systems are an integral part of many online services nowadays, particularly in e-commerce.
Amazon.com was one of the first organizations that relied on automated recommendations at scale, and early reports state that recommendations once accounted for more than 30\,\% of cross-sales on the site \citep{Linden2003}.
Today, recommender systems are found almost everywhere, and their potential value to businesses in terms of various key performance indicators has been demonstrated in many application settings \citep[cf.][]{jannachjugovactmis2019}.

Academic research in this area primarily focuses on \textit{consumer value}. The general idea is that recommender systems help users discover relevant content or support their decision-making processes by pre-selecting items from a more extensive catalog. These benefits of recommender systems are then assumed to translate to value for providers in the long run, e.g., to increased engagement and loyalty that finally translate into profit.

However, the mentioned two goals, i.e., maximizing consumer value and provider value, may represent a trade-off in specific applications. Furthermore, the strength of this trade-off may depend on the considered time horizon. An owner of an e-commerce shop, for example, might be tempted to recommend items that have a high profit margin instead of suggesting items that would be the best possible choice for the consumer. Many consumers will probably initially accept such suggestions, given the known persuasive potential of personalized recommendations \citep{PersuasiveRS2013}. Over time, however, they might lose their trust in the recommendations or the provider as a whole when they repeatedly find that the recommendations were not a good fit.

From an organizational perspective, the goal is to find a recommendation strategy that balances the two competing objectives. However, determining such a strategy cannot be done based on a static one-time investigation, as, for example, done for a numerical profit-relevance trade-off analysis in \cite{JannachAdomaviciusVAMS2017}. Instead, a longitudinal perspective must be taken, which considers repeated individual experiences of consumers over time to understand the long-term effects of a given strategy \citep{adomavicius2013understanding}.

In this work, we propose to rely on agent-based modeling (ABM) and simulation as a methodology to study the described longitudinal dynamics of recommender systems. ABM has been successfully applied in various other domains, but only a few works have explored its use to study the effects of recommendations on consumers and other stakeholders \citep{longitudinalimpact2021,consumptionPerformance2020,donkers2021dual}. In our approach, a recommendation provider and consumers are our simulation model's central elements (agents). Providers make recommendations based on a set of selected recommendation algorithms. These algorithms may consider either one or both dimensions, namely the profitability of the items and their assumed relevance to individual consumers. Consumers, in turn, may decide to accept a recommendation with a certain probability. This consumption probability depends on their trust in the provider---which is driven by their satisfaction with the recommendations---and may vary over time. Additionally, the consumption probability depends on the reputation of the provider on social media, which emerges as consumers share their experiences with the recommendation service.\footnote{\textcolor{black}{We emphasize that in our model, the consumers' decisions are independent of the profit margin for the provider, i.e., the consumer has no knowledge about profits, which is a realistic assumption in practice. Moreover, we make no assumptions regarding the relationship between end consumer \emph{price} and profit margin, i.e., higher-prized products may or may not have a higher profit margin than lower-priced ones.}
}

We ran various simulations to analyze the effects of different provider strategies on consumer trust and the resulting effects on profit. As the primary outcome, we found that purely or mainly provider-oriented recommendation strategies may lead to vastly increased overall profit compared to a purely-consumer-oriented strategy \textit{in the short run}. However, a recommendation strategy that balances consumer utility and profit---with more emphasis on the consumers' perspective---leads to a stable level of consumer satisfaction and sustained profitability \textit{in the long run}. Considering social media, furthermore, reinforces the observed results.

Overall, with our work, we aim to narrow the open research gaps in the area of multi-stakeholder recommender systems, which---despite their high practical importance---is still a relatively new research area. Given a lack of publicly available datasets, we address the problem in a novel simulation-based approach. Therefore, this work's main contribution lies in a novel ABM-based simulation model designed to study the longitudinal effects of different price- or consumer-oriented recommendation strategies. Furthermore, we ran extensive simulations highlighting the importance of balancing consumer value and provider profitability for sustained business success. The software framework developed for the simulations is, in principle, applicable in other application settings, particularly for experience goods \citep{hutter2011experience}. We publicly share the code of the simulation framework to foster future research.

The paper is organized as follows. Next, Section \ref{sec:related-work} provides more background on our methodology and discusses related works. Section \ref{sec:model} formalizes the \textcolor{black}{agent-based} model. Section \ref{sec:results} presents and discusses the results. \textcolor{black}{Section \ref{sec:conclusion} concludes our work and describes research limitations and possible extensions to our model.}

\section{Background and Related Work}
\label{sec:related-work}

In this section, we first provide the background on the important topic of multi-stakeholder recommendation and on agent-based modeling (Sec. \ref{subsec:background}). Afterward, we discuss existing simulation-based approaches from the recommender systems literature (Sec. \ref{subsec:simulation-in-recsys}).

\subsection{Background}
\label{subsec:background}

\paragraph{Multi-stakeholder recommendation and longitudinal effects}
Traditionally, research in the area of recommender systems focuses on consumer value (or consumer utility). In the vast majority of works, the research problem is operationalized as a machine learning problem, where the goal is to predict the relevance of items to individual users. The underlying assumption is that better algorithms would position the truly relevant items more prominently in a list of recommendations, thereby making search and discovery easier for consumers. Furthermore, it is often implicitly assumed that consumers will more often spot relevant items if they are provided in a recommendation list, which, as a result, will lead to consumption (e.g., in case of media recommendations) or purchase (e.g., in an e-commerce shop). However, the business value of these approaches is mostly not explicitly discussed in previous research.

While the assumption of an indirect relationship between recommendations and profit is undoubtedly reasonable, in many cases, it may represent a severe oversimplification of the problem that actual organizations face. First, in reality, not all items may be equal in terms of the value they create if they are successfully recommended to consumers. For example, as \cite{JannachAdomaviciusVAMS2017} discussed, different profit margins may be attached to different items, and recommending items with higher margins might be favorable from the provider side \citep{Hinz2010Impact}. Second, the key performance metrics that organizations collect for a deployed recommender system might not align or even conflict with the goal to recommend the most relevant items. News recommendations, for example, might be optimized to obtain a high click-through rate. However, such an optimization might lead to the effect of promoting click-bait items, which raise the consumers' attention and stimulate them to click on the item, only to detect afterward that the news was not relevant to them. Repeated bad experiences of this type might ultimately drive consumers away from the recommendation service or the online site as a whole in the long run. Finally, more stakeholders might be involved in the process than the recommendation provider and the consumer, such as manufacturers or providers of the items \citep{abdollahpouri2020,jannach2021mcnamara}.

Therefore, recent research indicates that it is essential to consider the perspectives of multiple stakeholders in the recommendation process and the longitudinal effects of recommendations \citep{adomavicius2021}. A vast body of research in the field of operations research is concerned with the trade-off between multiple objectives \citep[see, e.g., ][]{gutjahr2016stochastic,zajac2021objectives}. However, this line of research is mainly interested in identifying and evaluating Pareto optimal solutions on the efficient frontier in a mathematically constrained solution space \citep{thies2019operations}.
We, on the contrary, do not focus on developing optimal solutions but are rather interested in the effects of selected (and also commonly used) recommendation algorithms. These algorithms are indeed meant to handle multiple objectives; however, we are interested in the dynamics they unfold \textit{in the long run}. There are some previous works that analyze the longitudinal effects of selected methods to deal with two or more objectives simultaneously in an organizational context \citep[e.g.,][]{ethiraj2009,leitner2014}. However, these approaches mainly consider that decision-makers have limited access to information and mostly no memory, perfect foresight, and follow a steepest ascent hill-climbing algorithm to increase their utility \citep{Wall2020}. In the context of recommender systems, however, the situation is different: the consumers are not necessarily utility maximizers, also because some recommended goods might be experience goods that cannot be evaluated before consumption \citep{nelson1970}. Instead, their future decisions (e.g., to buy a product or not) depend, amongst others, on their experience (e.g., whether they were satisfied with previous recommendations or not), and information sharing between consumers, which promises interesting---and so far merely researched---dynamics.

We discuss selected previous \emph{simulation-based} works in the specific area of recommender systems below in Sec.~\ref{subsec:simulation-in-recsys}. \textcolor{black}{In our present work, we continue these recent lines of research, and we particularly focus on sustained profitability, which we see as a research gap that is not sufficiently explored yet} in previous research.

\paragraph{Agent-based modeling and simulation}
We rely on agent-based modeling as a research methodology because it allows us to efficiently study the longitudinal effects of different recommendation strategies in a bottom-up fashion: the modeling process takes place at the micro-level, i.e., at the level of individual consumers and recommendation providers, and makes it possible to gain insights into the macro-phenomena that emerge from micro-level interactions over a more extended period \citep{Epstein2006,Wall2020,Leitner2015}. Usually, ABMs consist of three building blocks that can be \enquote{designed} at the micro-level: \textit{(i)} autonomous and heterogeneous agents, \textit{(ii)} the environment in which agents operate, and \textit{(iii)} interactions among agents and among agents and the environment. A large number of degrees of freedom in designing these blocks allow for a rich contextualization, give exceptional control over framework conditions, and allow us to explore and isolate the effects of a multiplicity of factors in a precise way \citep{adomavicius2021,Wall2020,consumptionPerformance2020}.

The importance of \textcolor{black}{agent-based} modeling and simulation as a research methodology is growing in many domains, including social sciences \citep{gilbert2000,jager2021}, economics \citep{gurgone2018,steinbacher2021advances}, information systems \citep{haki2020}, and behavioral operations research \citep{leitner2017,robertson2016}. The use of \textcolor{black}{agent-based} modeling
and simulation in research on recommender systems has been increasing in recent years as well. These methodologies are particularly suited to study  longitudinal effects of recommendation algorithms for different reasons: \textit{(i)} they allow us to model a large population of consumers, with realistic characteristics, who individually react to recommendations provided by organizations,
\textit{(ii)} recommendations can be made according to  different recommendation strategies, and we can easily switch between strategies to isolate their effects, and
\textit{(iii)} observations can be made for more extended time. In contrast, due to prohibitively high costs, studying the  longitudinal effects of different recommendation strategies is very challenging when done with more traditional approaches such as field studies \citep{consumptionPerformance2020}.

\subsection{Simulation-based Research in Recommender Systems}
\label{subsec:simulation-in-recsys}
Simulation-based research of longitudinal dynamics,\footnote{Other types of simulations, e.g., of a user interacting with a conversational recommender system \citep{Greco2017Converse}, can also be found in the literature. However, such individual user simulations are not related to works that focus on longitudinal effects in a multi-agent environment.} as mentioned,
has obtained increased research interest in the community, as it becomes more and more evident that traditional ``one-shot'' evaluations of machine-learning models are too limited to assess several aspects of practical importance.

In one of the earlier works, \cite{FlederBlockbuster2009} developed an analytical model and found that recommender systems, over time, may \emph{decrease} sales diversity, which they measure in terms of the Gini index, and reinforce the popularity of already existing items.
Several subsequent works build on the ideas put forward in \cite{FlederBlockbuster2009}. For example, \cite{Hinz2010Impact} also look at the effects of recommender systems on sales. Specifically, they investigate to what extent the provision of search and recommendation functionality may lead to additional consumption and substitution effects. According to their analyses in the video-on-demand domain, recommendations may be particularly advantageous when they help to substitute low-margin blockbusters with high-margin niche products. Besides profitability, the authors also analyze concentration effects and the effects of providing non-personalized lists of the most popular items to users in different market situations. 
The more recent work by \cite{bountouridis2019siren} also builds on the early ideas by Fleder and Hosanagar. It proposes a simulation framework that can be used to study the longitudinal effects of different algorithms on long-tail novelty and unexpectedness in the news domain. The authors also consider the phenomenon of user interest drift in their work, and they rely on several additional metrics, e.g., to measure diversity.

The effects of different algorithms on sales concentration and popularity reinforcement over time were also discussed more recently  for rating prediction and ranking problems \citep{JannachLercheEtAl2015} and session-based recommendation problems \citep{ferrarojannachserra2020recsys}. While the technical approach differed from the analytical one in \cite{FlederBlockbuster2009}, these works led to similar insights. Both in \cite{JannachLercheEtAl2015} and \cite{ferrarojannachserra2020recsys}, it turned out that the choice of the algorithm matters and that some algorithms have a more substantial popularity bias than others. Furthermore, it was found that some algorithms exhibit lower levels of personalization and, therefore, tend to recommend a small set of items to everyone, which finally leads to possibly undesired concentration effects.

A few works in the related literature use simulation to study the performance of algorithms. In \cite{Umeda2014EvaluationOC}, for example, the authors present the results of an exploratory study. They rely on a simulation procedure to understand the effectiveness of nearest-neighbor algorithms and suitable parameter choices. More recently, \cite{consumptionPerformance2020} used agent-based modeling to study the performance development of recommendation algorithms over time. In their simulations, they observed an unexpected phenomenon which they called the ``performance paradox''. Specifically, they found that if consumers rely strongly on the provided recommendations, the system's recommendation quality will decrease over time in terms of prediction accuracy and diversity. Furthermore, additional simulations indicated that using a hybrid approach that combines personalized and non-personalized recommendations can help avoid this paradox.

Building on the same ABM framework, \cite{longitudinalimpact2021} investigated the potential problem of \emph{preference biases}. Such an effect may occur when consumers do not provide their \emph{true}, unbiased assessments (ratings) of items to the system but a rating influenced by the recommendation system itself. This, in particular, happens when the recommender system not only shows items to users to watch but also presents some average community rating or a predicted rating along with the recommendation. Ultimately, this may lead to ``polluted'' (noisy) preference information, which propagates through the system over time.

The work presented in this paper is related to previous work in that it, like \cite{consumptionPerformance2020} and \cite{longitudinalimpact2021}, builds on the ABM framework as a research methodology due to its potential advantages as described in Sec.~\ref{subsec:background}. Moreover, we also consider questions related to the profitability of individual items as discussed, e.g., in \cite{Hinz2010Impact}. The goals of our research are, however, different. Whereas previous works often focus on diversity and popularity reinforcement questions, we aim to provide a mechanism for decision support for service providers who seek balanced recommendation strategies that ultimately lead to sustained business success. In our work, we report the outcomes for a selected application domain. However, the employed methodology is by no means tied to a particular domain.

While our work addresses a rather general problem, there are also works in the literature that employ different types of (non-ABM) simulation techniques to study specific application-specific issues, e.g., in the context of technology-enhanced learning \citep{SIE20102883,Nadolski2009}, where the examined problems, for example, include coalition formation in learning situations or the prediction of dropouts. Furthermore, \cite{prawesh2014most} use a  simulation-based approach to examine the manipulation resistance of non-personalized ``most popular items'' recommendations in the news domain. However, the relation of our work to these research efforts is marginal.

Finally, in the most recent years, some simulation-based approaches were put forward that have the goal to provide environments in which recommendation algorithms based on reinforcement learning (RL) can be evaluated and benchmarked. These simulation-based platforms, in particular, include RecSim \citep{ie2019recsim}, RecSim NG \citep{mladenov2021recsimng}, SOFA \citep{huang2020keeping}, RecoGym \citep{rohde2018recogym}, and PyRecGym \citep{shi2019pyrecgym}. One main goal of these approaches is to evaluate RL-based strategies in a realistic environment without conducting complex and expensive field tests. While these frameworks also share with our work the goal of analyzing the effects over time, the general relation to our work is limited. Specifically, these frameworks focus on the performance evaluation of a particular machine learning method, while we aim to model the complex behavior of heterogeneous and independently acting agents.

\subsection{Summary of Relation with Previous Work and Research Gaps}
\label{subsec:relation-and-gap}

Recommender systems research has historically focused on consumer value and accurate preference predictions at a given time. Our work, in contrast, continues very recent lines of research, which \emph{(a)} acknowledge that recommendation is a problem that requires us to consider multiple stakeholder goals in parallel and \emph{(b)} emphasize the importance of assessing longitudinal effects. Our review of related research revealed that a few works exist in the literature which focus on questions of (economic) effects of recommender systems, but to our knowledge, none of them aims to assess the performance of different provider strategies through a simulation from a longitudinal perspective. With our present work, we aim to narrow this research gap by proposing a simulation approach based on agent-based modeling, a technique that was recently also explored to study longitudinal dynamics of recommender systems \citep{consumptionPerformance2020}, although not from a multi\textcolor{black}{-}stakeholder and value-oriented perspective.

\section{Proposed Simulation Model}
\label{sec:model}

\subsection{Model Overview}
\label{sec:simulation-model-overview}

\begin{figure}[h!t]
\centering
\includegraphics[width=.8\linewidth]{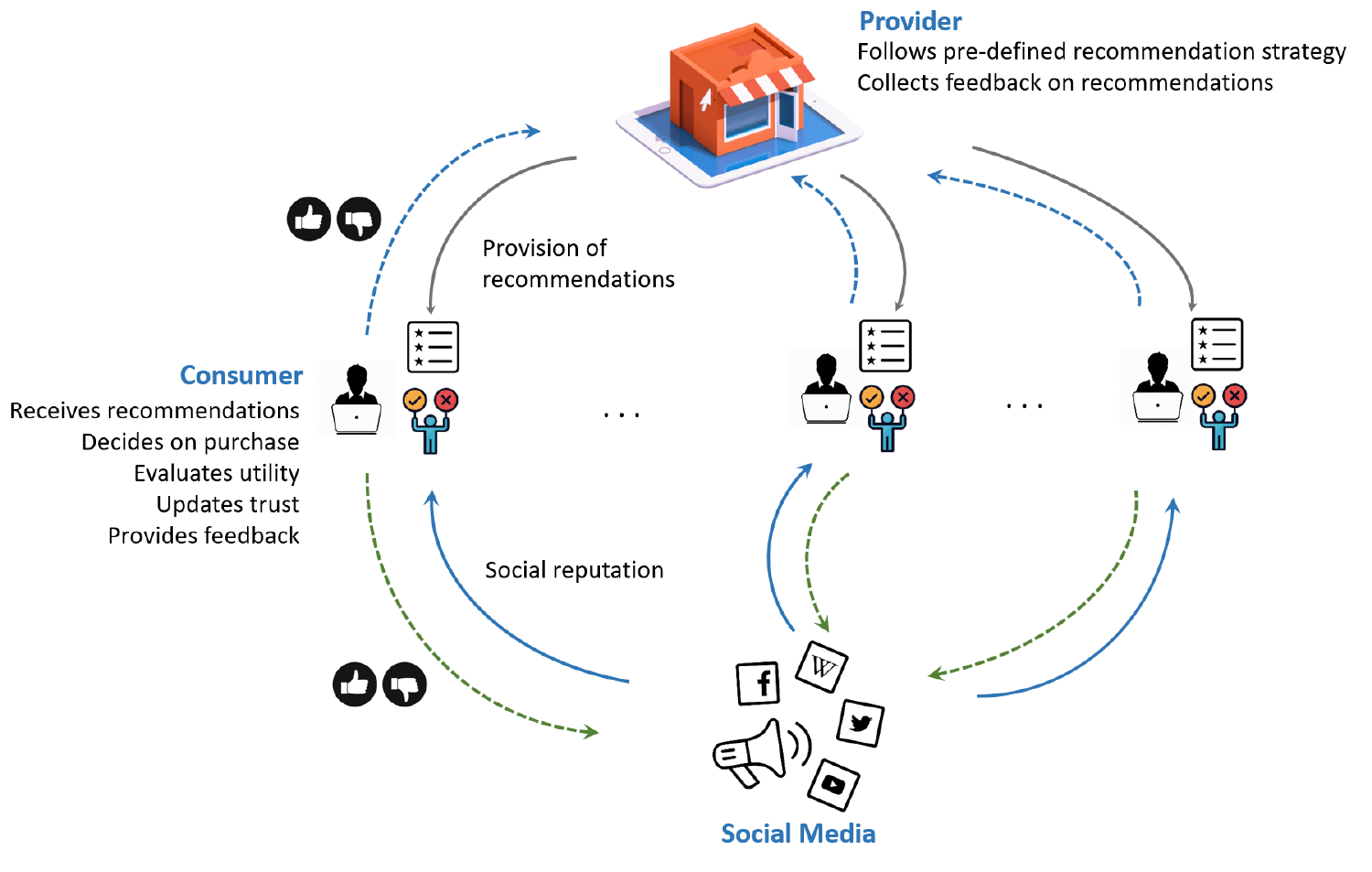}
\caption{High-level overview of the simulation model.}
\label{fig:big-picture}
\end{figure}

The main goal of our simulation, as mentioned above, is to study the longitudinal effects of selected recommendation strategies on consumer trust, consumption probability, and, consequently, on long-term business success. Figure~\ref{fig:big-picture} provides a high-level overview of the model.

Central to our model is an organization (e.g., an e-commerce business), i.e., a provider that offers goods to consumers and provides (personalized) recommendations as an additional service. The provider relies on one specific strategy (or algorithm) to generate the recommendations. In our simulations, we explore the effects of several such strategies, including a strategy based on the popularity of goods, one that aims to maximize consumer utility, one that seeks to maximize the provider's \textcolor{black}{profit}, and strategies that balance the possibly competing goals. \textcolor{black}{Please note that by profit we refer to the financial gain of the provider (i.e., revenues minus costs) and not to a price paid by the consumer whenever an item is consumed.} As usual in modern recommender systems, the provider relies on consumer feedback about items to personalize the recommendations \citep{JannachZankerEtAl2010}. \textcolor{black}{In our simulation, we assume that consumers provide \emph{explicit} feedback in the form of numerical ratings.}

The other \emph{agents} in our system are the consumers, who are the recipients of the recommendations. We assume that they accept individual recommendations by the provider with some probability, i.e., they purchase one of the recommended items from time to time. The most important aspect to note in this context is that the probability of buying a recommended item depends on two factors: \emph{(i)} the agent's \emph{own past experiences} (that capture whether previous recommendations were considered valuable or not) and \emph{(ii)} the \emph{social reputation} of the provider. \textcolor{black}{Please also note that we model cases in which consumers do not consider the price of goods when making consumption decisions. In some cases, consumers might not even be aware of the individual prices of the goods they consume, which might, for example, be the case when recommendation providers offer flat-rate subscription models, which is often the case in the media streaming domain.}
Whenever consumers accept a recommendation and purchase, they evaluate the experienced utility against their expectations (often referred to as preference match)
and record the outcome as their own experience with the provider. Please note that the provider offers experience goods only. Consequently, consumers can only evaluate the utility \textit{after} consuming an item \citep{nelson1970}. From time to time, consumers also provide \textcolor{black}{numerical} feedback about their experience to the provider, and the provider updates the user-specific preference profiles at \textcolor{black}{given} intervals. Sometimes, consumers also post their experiences with the recommendations to a social media channel. \textcolor{black}{This can, for example, take the form of a thumbs up/down rating of the provider on a social media platform.} \textcolor{black}{Finally,} the aggregated feedback of all consumers determines the provider's social reputation. Algorithm \ref{alg:simulation-loop} sketches the main simulation loop of our approach.

Overall, one expectation is that focusing too much on the provider's profit will lead to repeated bad experiences on the consumers' side because they frequently receive recommendations that do not match their preferences well. Consequently, the consumers' trust in the provider may decrease over time, which might lead to a lower probability of purchasing one of the recommended items in the future. Furthermore, when the negative experiences are shared on social media, the social reputation of the provider will perhaps drop, as consumers might become more and more unsatisfied, and the bad reputation may influence the purchase probabilities of other agents as well. In consequence, the overall profit of the provider may decrease (more or less slowly) if a recommendation strategy is employed that focuses too much or exclusively on \textcolor{black}{the provider's} (short-term) profitability.

\begin{center}
\begin{minipage}{0.85\textwidth}
\begin{algorithm}[H]
\caption{Main simulation loop (sketch)}\label{alg:simulation-loop}
\begin{algorithmic}[1]
\State \emph{Initialize simulation model}: Consumer preference profiles and recommendation strategy
\For{specified number of iterations}
    \State Provider delivers recommendations to consumers
    \For{each Consumer agent}
        \State \parbox[t]{295pt}{Consumer decides on acceptance of one recommendation based on own experiences and social reputation of the provider\vspace{6pt}\strut}
        \If{Consumer accepts one recommendation}
            \State \parbox[t]{295pt}{Consumer evaluates utility of recommended item and updates
            own past experiences\vspace{6pt}\strut}
            \State Consumer posts experience to provider with certain probability
            \State Consumer posts experience to social media with certain probability
        \EndIf
    \EndFor
     \State Provider updates preference profiles on regular intervals
\EndFor
\end{algorithmic}
\end{algorithm}
\end{minipage}
\end{center}

\subsection{Model Details}
\label{sec:simulation-model-details}

The simulation was developed in Python 3 using the Mesa modeling framework \citep{python-mesa-2020}. We share all code and data \textcolor{black}{online}.\footnote{The code is available at \url{https://github.com/nadaa/rec-strategies-abm/}; the simulation data can be found here: \url{https://seafile.aau.at/f/1e84eec700244ac1871d/}}
Instructions for installing and reproducing the results are available in the provided documentation as well.

\subsubsection{Preliminaries and Model Initialization}
We model a system populated by one recommendation provider and $C \in \mathbb{N}$ consumers. The provider offers a total of $M\in \mathbb{N}$ items to consumers. Our model considers a transaction-based revenue model for the provider, which is a prevalent approach in e-commerce.\footnote{Alternative models could, for example, be subscription-based.} As an application domain, we consider the case of movie recommendations \citep{MovieLens2016,bennett2007netflix,Ekstrand:2014:UPD:2645710.2645737},
which is the focus of many previous research works. According to the assumed provider revenue model, consumers pay per stream or download, as usually done also on commercial services like Amazon Prime Video.

As commonly done in previous research, we assume that recommendations only include items that the user has not purchased before. This is usually done to support the discovery of new items. Following the approach in \cite{longitudinalimpact2021}, we furthermore fix the item catalog during the simulations. Dynamically adding and removing items is easily possible in our model. However, in the present work, we focus on the main aspects of interest, namely the development of the consumption probability and the provider's profit, given that different recommendation strategies are employed.

The provider agent knows the consumer preferences. This is a feasible assumption as the provider collects them through an online service platform. \textcolor{black}{In our model, we assume that consumers provide feedback in the form of a numerical rating.} \textcolor{black}{Then,} to predict the utility of an item that an individual user has \textit{not} purchased before, the provider employs a machine learning model based on matrix factorization \citep{koren2008factorization}, which is an approach that is also widely used in practice. Therefore, the recommendations forwarded to individual users are based on the rating predictions of the machine learning models.

Our approach is generically applicable to transaction-based revenue models and not tied to a particular application domain. As mentioned, our simulations use the movie domain as an example domain. We initialize the consumer agents in our model based on movie ratings provided by users of the MovieLens online recommendation service.\footnote{The MovieLens dataset \citep{movielens} contains 100k ratings by 610 MovieLens users for 9,742 movies. Consequently, our simulation model is populated with 610 consumers, each initialized with a preference profile based on the MovieLens dataset. Other datasets with rating (preference) information could be used as well.} We note that the MovieLens dataset does not contain information about the profitability of individual movies for the provider, which is why we generate synthetic profit data.
We assume that not all movies lead to the same profit, which is a feasible assumption since some movies may have a higher profit margin than others, e.g., because of different royalties that must be paid. As also done by \cite{JannachAdomaviciusVAMS2017} and \cite{abdollahpouri2020}, we draw the synthetic profit data from a Normal distribution.

\subsubsection{Item catalog}
\label{sec:item-catalog}
The provider offers items $m_i$ to consumers, where $i=1,\dots,M$. We denote the fixed set of items by $\mathbf{M}$. To provide personalized recommendations, we keep track of every consumer's consumption history. In particular, we denote the set of items consumed by consumers $c_j$ up to period $t$ by $\mathbf{M}_{jt}$. We intend not to recommend items that were already consumed before. Therefore, we denote the set of items that can potentially be recommended to consumer $c_j$ in period $t$ by $\mathbf{\overline{M}}_{jt}$, and define it as the complementary set of $\mathbf{M}_{jt}$ in $\mathbf{M}$.

All items are characterized along three dimensions.
For item $m_i$, these dimensions are
\textit{(i)} the \textcolor{black}{numerical} feedback $\mathbf{R}_{i}$ that the organization has received from consumers after they have consumed the item,
\textit{(ii)} the utility $\hat{\mathbf{R}}_{it}$ that the organization predicts in $t$ (based on the received feedback) to materialize after consuming the item,
\textit{(iii)} the profit $\rho_i$ related to item $m_i$:

\begin{equation}
m_i = \{\mathbf{R}_{i}, \hat{\mathbf{R}}_{it},\rho_i\}~.
\end{equation}

\paragraph{Consumer feedback}
Over time, consumers buy items and might provide \textcolor{black}{explicit feedback} on the utility they experienced from the consumption. We denote the feedback as \textcolor{black}{numerical} ratings that the organization has received on item $m_i$ by $\mathbf{R}_{i} = (r_{i1},\dots,r_{iC})$. Please note that not all consumers necessarily provide feedback on the items they consume, which is why there might be missing entries in $\mathbf{R}_{i}$. In particular, if consumer $c_j$ has already provided feedback on item $m_i$, then $r_{ij} \in \mathbb{R}$ and $r_{ij} \in \left[0,5\right]$. In contrast, if consumer $c_j$ did \textit{not yet} provide feedback, $r_{ij}=\{\}$. The initial feedback on good $m_i$ is taken from the MovieLens dataset, and $\mathbf{R}_{i}$ is continuously updated during the simulation.

\paragraph{Predicted utility}
To optimize the recommendation service, the provider uses the \textcolor{black}{numerical} feedback provided by consumers and predicts the missing values in $\mathbf{R}_{i}$. The predicted utilities of any consumer and item are a function of the previous ratings of a set of consumers and items \citep{adomavicius2005toward}. Technically, we use a collaborative filtering approach based on matrix factorization, as mentioned above, to implement the prediction function.\footnote{We use the SVD class from the \emph{surprise} recommendation library, with latent factors set to 100 \citep{Hug2020}.}
We refer to the set of actual feedback on item $m_i$ and the predictions of the missing values in $t$ by $\hat{\mathbf{R}}_{it} = (\hat{r}_{i1t},\dots,\hat{r}_{iCt})$. Please note that there are no missing values in $\hat{\mathbf{R}}_{it}$; in consequence, all $\hat{r}_{ijt} \in \left[0,5\right] \in \mathbb{R}$. The interval is the rating scale that is used in the underlying MovieLens dataset.\footnote{Since the database of ratings is extended during the simulations, we re-compute the threshold value when we re-compute the rating predictions, i.e., after 100 time steps.}

\paragraph{\textcolor{black}{Synthetic profit data}}
In addition to the feedback given by consumers and the predicted utilities, an item is characterized by the profit the organization can generate when \textcolor{black}{the item} is consumed. \textcolor{black}{Please recall that the profit does not reflect the price the consumer pays but stands for the financial gain (i.e., revenue minus costs) the provider makes when an item is consumed.} In particular, for every item  $m_i$, we draw the profit $\rho_i$ from a Normal distribution so that $\rho_i \sim N(2.5,1)$, with lower and upper bounds of 0 and 5, respectively. Please note that we model digital items that can be consumed by multiple consumers simultaneously. \textcolor{black}{We iterate that profit information is only available to providers, but not to consumers.}

\subsubsection{Provider Behavior}
\label{sec:provider-behavior}
The task of the provider agent is to provide $\mathcal{N}$ recommendations to consumers following a pre-defined recommendation strategy. Furthermore, the provider may receive feedback from consumers about items during the simulation. This feedback is considered in the computation of the predicted utilities $\hat{\mathbf{R}}_{it}$, consequently affecting subsequent recommendations. Finally, the provider keeps track of the purchases and the profit of each purchase. Note that the overall profit that the provider accumulates is a central output variable of interest.

\paragraph{Recommendation strategies} We consider recommendation strategies that either consider the consumers' predicted utilities and the provider's profit or are based on item popularity. To describe the former class of strategy, we introduce the weighting factors $\omega$ and $1-\omega$ to weigh the predicted consumer utility ($\hat{r}_{ijt}$) and the related profit ($\rho_i$), respectively. We consider the following strategies:

\begin{itemize}
\item \textbf{Consumer-centric}: this is the most common approach in the literature. It consists of recommending the items with the highest predicted utilities to consumers, and, consequently, aims to maximize the consumers' satisfaction with the recommendations \citep{adomavicius2005toward}. In particular, to compute the recommendations for consumer $c_j$ in period $t$, the provider sorts the items $m_i \in \mathbf{\overline{M}}_{jt}$ in descending order based on the related predicted ratings $\hat{r}_{ijt}$. \textcolor{black}{Recall that $\mathbf{\overline{M}}_{jt}$ indicates the items that can be recommended to consumer $c_j$ in period $t$, i.e, those items the customer has not yet consumed.} We denote the sorted list by $(\mathbf{\overline{M}}_{jt}, \geq, \hat{r})$. The recommendation list for consumer $c_j$ comprises the top $\mathcal{N}$ items of the ordered list $(\mathbf{\overline{M}}_{jt}, \geq, \hat{r})$. Since the provider only considers the consumers' utilities and does not take into account profit when preparing recommendations, $\omega$ is equal to $1$ in this case.

\item \textbf{Profit-centric}: in this strategy, the provider recommends the items \textcolor{black}{characterized by} the highest profit to consumers. This strategy is not personalized, except that already purchased items are not recommended again to a user. Let us denote the list for consumer $c_j$ in period $t$ that is sorted by profit by $(\mathbf{\overline{M}}_{jt}, \geq, \rho)$. Consumer $c_j$'s recommendation list comprises the top $\mathcal{N}$ items from that list. When following this strategy, the provider only considers the profit when preparing recommendations. Thus, $\omega$ is equal to $0$ in this case.

\item \textbf{Balanced and consumer-biased}: the provider aims to achieve a balance of utility and profit orientation in these strategies. We denote the list sorted by this order criterion by $(\mathbf{\Bar{M}}_{jt}, \geq, \omega \cdot \hat{r} + (1-\omega) \cdot \rho)$,
and use the top $\mathcal{N}$ items from that list as recommendations for consumer $c_j$. For the balanced strategy, we set $\omega = 0.5$, and for the consumer-biased strategy, we fix $\omega$ at $0.9$.

\item \textbf{Popularity-based}: finally, we consider the non-personalized recommendation of generally popular items as a baseline in our simulations. Recommending popular items is often considered a simple, \enquote{save}, and yet often effective strategy in practice. In our experiments, we use the number of existing ratings for each item in the dataset as an indicator for the popularity of an item.
Let us use the number of ratings an item has received as the order criterion for the popularity-based strategy. We formalize the function to count the received ratings of item $m_i$ by $ \sum_{j=1}^{J} \left[r_{ij} \neq \{\}\right]$. Following the Iverson bracket notation, the logical proposition $P$ in the brackets is either $\left[P\right]=1$, if the proposition is satisfied, i.e., when consumer $c_j$ already submitted a rating for item $m_i$, or $\left[P\right]=0$, otherwise \citep{Graham1989}. We denote the sorted list by $(\mathbf{\overline{}{M}}_{jt}, \geq,\sum_{j=1}^{J} \left[r_{ij} \neq \{\}\right])$ and forward the top $\mathcal{N}$ items from that list as recommendations to consumers. Again, we avoid that already consumed items are on that list.
\end{itemize}

\paragraph{Feedback collection and retraining} During the simulation, the provider is informed when a consumer selects an item for \textcolor{black}{consumption} and when a consumer provides rating feedback for a \textcolor{black}{consumed} item. Note that consumers do not give rating feedback for all their purchases. The provider collects the \textcolor{black}{numerical} feedback and adds it to the feedback database $\mathbf{R}_i$. At regular intervals---after $100$ periods in our experiments---the predicted utilities $\hat{\mathbf{R}}_{it}$ are recomputed. Therefore, the recommendations for each consumer change over time due to the individual feedback that a consumer provided, the collective feedback of all consumers, and the consumption of items (since items are not recommended again to consumers once they consume them).

\subsubsection{Consumer Behavior}
\label{sec:consumer-behavior}
Upon receiving recommendations, a consumer can make several choices. Note that the provider's behavior is mainly deterministic,\footnote{Except for the randomized initialization of the latent factors in the learning process.} whereas several stochastic elements drive the behavior of consumers. \textcolor{black}{Every consumption is decided upon in a two step process: First, the decision to consume an item} is affected by the individual \emph{consumption probability}; this probability might depend on both a consumer's experience with a recommendation provider and the current social reputation of that provider. \textcolor{black}{Second, if a consumer decides to consume an item, they select one of the $\mathcal{N}$ recommended items based on a probability distribution.} \textcolor{black}{In the following paragraphs, we first explain how we model consumer trust and the provider's social reputation. These measures are, then, aggregated to a consumer-specific consumption probability. Finally, we introduce the model according to which consumers select items they want to consume from a list of recommended items. }

\paragraph{\textcolor{black}{Consumer satisfaction and the formation of trust}}
A central question in \textcolor{black}{our research} context is how to model a consumer's experience \textcolor{black}{with a recommendation provider, and how individual consumer trust emerges from previous experiences}. In particular, it needs to be defined whether or not a consumer is satisfied with a recommendation, or, in other words, if the experience with a recommendation was a positive or a negative one. Recall, in period $t$, items $m_i \in \mathbf{\overline{M}}_{jt}$ can potentially be recommended to consumer $c_j$. We model consumers who expect the provider to recommend only those items out of $\mathbf{\overline{M}}_{jt}$ that best meet their preferences, and operationalize this expectation by introducing a consumer expectation quantile $\psi$ that translates into an individual expectation threshold (see Eq. \ref{eq:threshold}). This approach to model customer satisfaction is in line with the expectation-confirmation theory \citep{oliver1980cognitive,bhattacherjee2001understanding}.
\textcolor{black}{In particular, we model that consumers are satisfied with the recommendation service if} they pick one of the recommended items, consume it and experience a utility that \textit{exceeds} their expectation threshold. \textcolor{black}{In other words, consumers are satisfied with the recommendation service if the items they consume after a recommendation are in the top $1-\psi$ quantile of all items (ordered by consumer utility) available to them}. In contrast, they are dissatisfied if their utility is below this threshold \citep{kim2009trust}.

\textcolor{black}{Recall that we only recommend items to consumers that were not consumed before by that customer. Thus, the set of items that can be recommended to consumer $c_j$ is dynamic. In consequence, the expectation thresholds are dynamic, too.} Let $f^{\hat{r}}_{jt}(m)$ be the probability distribution function of the predicted utilities $\hat{r}_{ijt}$ of items $m_i \in \mathbf{\overline{M}}_{jt}$, \textcolor{black}{i.e., of those items that can potentially be recommended to consumer $c_j$ in period $t$}. Then, consumer $c_j$'s individual expectation threshold in period $t$, $\underline{u}_{jt}$, is \textcolor{black}{the utility of the item at the $\psi$ quantile}, i.e.
\begin{equation}
\label{eq:threshold}
    \int_{0}^{\underline{u}_{jt}} f^{\hat{r}}_{jt}(m)~dm = \psi~.
\end{equation}
\noindent\textcolor{black}{This means that $1-\psi$\,\% of all items that can be recommended to consumer $c_j$ in period $t$ have a predicted utility higher than $\underline{u}_{jt}$, and consumers are satisfied if they get recommended and consume one of those particular items.} Please note that consumers are homogeneous with respect to the expectation quantile $\psi$. However, since consumers are heterogeneous with respect to the predicted utilities and because the resulting expectation threshold is a function of the predicted ratings related to goods in $\mathbf{\overline{M}}_{jt}$, consumers are heterogeneous with respect to the expectation threshold $\underline{u}_{jt}$.

Consumers experience utility whenever they consume an item. Given the vast success of collaborative filtering approaches in recommender systems in practice \citep{singh2020collaborative}, we assume that the ratings predicted with the matrix factorization model are a good approximation of the \textcolor{black}{consumers'} true utilities. Since such predictions are, of course, not exact, we define the true utilities as the predicted rating plus some noise:
\begin{equation}
u_{jt} = \hat{r}_{ijt} + \epsilon~,
\end{equation}
where $\epsilon \sim N(0,0.3)$.\footnote{Consumers pick at most one item from the list of recommended items per time step. For the sake of readability, we omit the notion of $i$ in the utility.}

After consuming an item, consumers recompute their trust in the organization by belief updating \textcolor{black}{using the mean of the Beta distribution, as proposed in prior literature as a method to compute trust \citep{liu2014computational}}. In particular, consumer $c_j$'s trust in the organization in period $t$ is based on $\{\alpha_{jt}, \beta_{jt}\}$\textcolor{black}{, whereby $\alpha_{jt}$ indicates the positive and $\beta_{jt}$ stands for the negative} experiences \textcolor{black}{of consumer $c_j$} with the provided recommendations up to that period.

We implement the \textcolor{black}{update of trust} as follows: after consuming item $m_i$ in period $t$, consumer $c_j$ updates $\{\alpha_{jt}, \beta_{jt}\}$ such that $\alpha_{jt}$ ($\beta_{jt}$) increases if the experienced utility exceeds (is below) the expectation threshold introduced in Eq. \ref{eq:threshold}. \textcolor{black}{This means that we increase $\alpha_{jt}$ and do not update $\beta_{jt}$ if consumers are satisfied with a recommendation, and vice-versa if consumers are not satisfied with the recommendation service.}
In particular, we use the Euclidean Distance between \textcolor{black}{consumer $c_j$}'s expectation threshold and the actual utility of that consumer in period $t$ for this update so that

\begin{equation}
\label{eq:trust-update}
 \{\alpha_{jt}, \beta_{jt}\} =
    \begin{cases}
    \{\alpha_{jt-1}+(u_{ijt}-\underline{u}_{jt})^2, \beta_{jt-1}\} &\text{if } u_{ijt} \geq \underline{u}_{jt}\\
    \{\alpha_{jt-1}, \beta_{jt-1}+(u_{ijt}-\underline{u}_{jt})^2\} &\text{if }u_{ijt} < \underline{u}_{jt}\\
    \end{cases}
\end{equation}

\noindent \textcolor{black}{The first case describes situations in which consumers are satisfied with the recommendations, whereas the second case captures dissatisfied consumers. We decided to use the Euclidean Distance for this update as this measure puts more (less) emphasis on large (small) deviations. In consequence, if consumers are provided with recommendation that exceed or undercut the consumers' expectations by large  numbers, this will have stronger effects as compared to recommendations that are relatively close to the consumers' expectations.} Finally, consumer $c_j$'s trust in the organization in period $t$ is \textcolor{black}{the fraction of consumer $c_j$'s positive and negative experiences with the provider up to period $t$:}

\begin{equation}
\label{eq:trust}
   c_{jt}^{\text{trust}} = \frac{\alpha_{jt}}{\alpha_{jt}+\beta_{jt}}~.
\end{equation}

With probability $p^{\text{feed}}$, consumers give feedback about the utility experienced after consuming an item, $u_{jt}$, to the provider.
If consumer $c_j$ provides feedback on item $m_i$ in period $t$, the provider updates the characteristics of item $m_i$ such that $r_{ij} = u_{jt}$.

\textcolor{black}{To initialize the consumers' experiences with the recommendation provider, we use a dataset of past user experiences.
Recall that the feedback database $\mathbf{R}_i$ contains only the rating data included in the dataset used for initialization in period $t=0$. We use this data to compute the initial values of $\alpha_{j0}$ and $\beta_{j0}$ according to
\begin{equation}
\label{eq:initial-alpha}
    \alpha_{j0} = \sum_{i=1} ^ I \left [ r_{ij0} \ge \underline{u}_{j0} \right ]
\end{equation}
\noindent and
\begin{equation}
\label{eq:initial-beta}
    \beta_{j0} = \sum_{i=1} ^ I \left [ r_{ij0} < \underline{u}_{j0} \right ]~,
\end{equation}}
\textcolor{black}{where $r_{ijo}$ is the initial rating included in the MovieLens dataset of consumer $c_j$ on item $m_i$. For simplicity, we assume that the initial expectation threshold is the mean of a consumer's ratings, and we compute it by $\underline{u}_{j0} = \sum_{i=1}^{M}r_{ij0} / \sum_{i=1}^{M} [r_{ij0} \neq \{\}]$, where the numerator is the sum of numerical feedback of consumer $c_j$ included in the dataset used for initialization, and the denominator is the number of items for which consumer $c_j$ provided feedback. Equations \ref{eq:initial-alpha} and \ref{eq:initial-beta} follow the Iverson bracket notation and are count functions: whenever an initial rating is above or equal (below) the threshold $u_{j0}$, the values of $\alpha_{j0}$ ($\beta_{j0}$) increases by one. The initial trust is computed following Eq.~\ref{eq:trust-update}, i.e., consumer $c_j$'s initial trust $c_{j0}^{\text{trust}}$ is $\alpha_{j0} / (\alpha_{j0} + \beta_{j0})$}.

\paragraph{\textcolor{black}{Sharing information on social media: The provider's social reputation}}
In addition to a consumer's own past experience, we model the provider's social reputation as a factor that can affect the consumption probability {of a consumer}. Specifically, a consumer may either post a corresponding \emph{thumbs-up} or \emph{thumbs-down} on a social platform. To that purpose, we assume that consumers post their experience with the provider on a social media channel\textcolor{black}{, and the decision of whether to rate the recommendation provider on social media (and to thereby affect the provider's social reputation) or not is driven by the consumer's past experience with the recommendation service}. Technically, the decision \textcolor{black}{of posting} on social media is based on a U-shaped Beta distribution\footnote{Specifically, we use the inverse of the Beta Distribution $B(2,2)$.}, which implements the intuition that the probability of posting to social media is higher for more \emph{extreme} (i.e., very good and very poor) experiences.\footnote{\textcolor{black}{The models of information diffusion in social networks employed by previous research often consider fixed information sharing probabilities \citep{dinh2020}. By doing so, these models ignore the agents' local information, which is often regarded as a limitation of previous models \citep{chen2021}. We, in contrast, explicitly consider the agents' local information and make the decision of whether to share information on social media contingent on an agent's experience with the recommendation provider.}}

The global social reputation of a provider is then the ratio of positive information shared on social media to all related social media posts. As mentioned, we consider binary data (e.g., shared via a thumbs-up and thumbs-down functionality) and denote the number of positive and negative information shared by all consumers up to period $t$ by $s^{+}_t$ and $s^{-}_t$, respectively. Finally, we formalize the provider's social reputation in period $t$ by

\begin{equation}
\label{eq:organization-reputation}
    c_{jt}^{\text{soc}} = \frac{s^{+}_t}{s^{+}_t + s^{-}_t}~.
\end{equation}

\noindent Please note that consumers do not \emph{update their knowledge of the provider's} social reputation in every period, but do so with probability $p^{\text{soc}}$\footnote{ In our model, we set $p^{\text{soc}}$=0.15.}. This means, if consumer $c_j$ updates the social reputation in a period, they do so following Eq.~\ref{eq:organization-reputation} and store $ c_{jt}^{\text{soc}}$ it in their memory. If they do \textit{not} update the social reputation in a period, $c_{jt}^{\text{soc}} = c_{jt-1}^{\text{soc}}$. Note that we propose to use a simple yet intuitive form of deriving a global reputation score from the observed social media posts based on the fraction of positive posts compared to all posts. Alternative schemes, which, e.g., weigh negative posts stronger than positive ones or which consider more fine-grained consumer feedback, are of course possible. In our simulation, we do not rely on an existing dataset to initialize the social reputation of the provider. Instead, the simulation starts with a neutral reputation, and the provider's reputation then develops based on the feedback by the agents, which in turn is determined with their quality assessment of the recommendations.

\begin{figure}
\centering
  \includegraphics[width=0.6\textwidth]{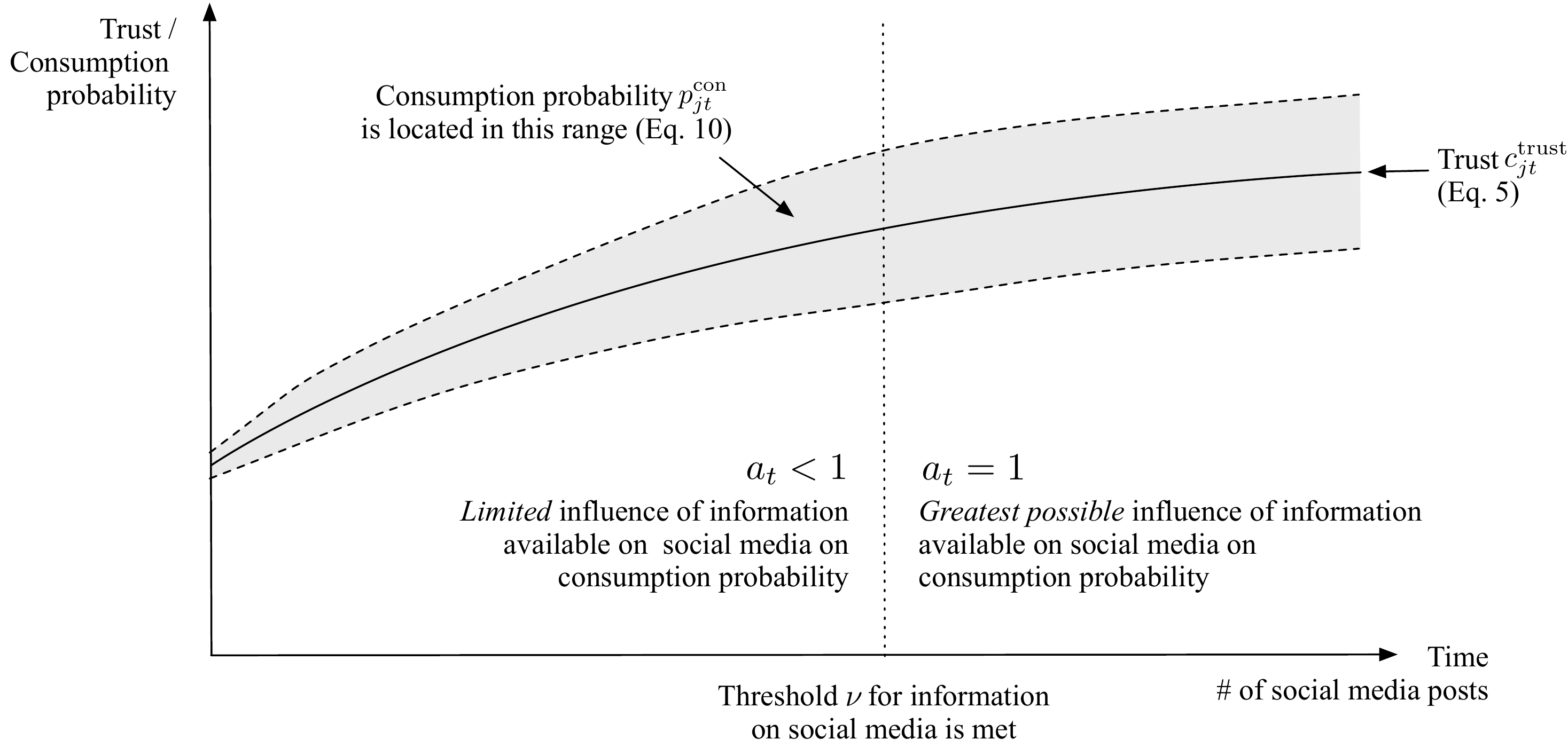}
  \caption{Consumption probability}
  \label{fig:consumption-probability}
\end{figure}

\paragraph{\textcolor{black}{Consumption probability: Combining own experiences with social reputation}}
Having defined consumer $c_j$'s trust in the recommendation provider, $c_{jt}^{\text{trust}}$, and the provider's reputation in social media, $c_{jt}^{\text{soc}}$, we can finally compute consumer $c_j$'s overall consumption probability $p_{jt}^{\text{con}}$. To do so, we use the individual trust $c_{jt}^{\text{trust}}$ as a reference point and limit the consumption probability to the range $c_{jt}^{\text{trust}} \pm \Delta$. The higher $\Delta$, the stronger the effect of social media on the consumers' consumption decisions. To additionally account for the \textit{volume} of information available on social media, we introduce the weighting factor
\begin{equation}
a_t = \min\left(\frac{s_t^+ + s_t^-}{\nu}, 1\right)~,
\end{equation}
where $\nu \in \mathbb{N}$ indicates the volume of information on social media required so that consumers fully take this information into account when making their consumption decisions. The rationale behind this modeling approach is that the consumption probability should not be over-proportionally affected by only a few posts on social media, which will be the case at the beginning of the simulations. Information on social media will be fully taken into account as soon as $\nu$ posts are available, and in the case of less than $\nu$ posts, the weighting factor regulates the impact of social media. Since the model is constructed so that social media posts are accumulated over time, its impact on consumption increases with $t$ up to a maximum value of $\pm \Delta$. The logic underlying the computation of the consumption probability is illustrated in Fig. \ref{fig:consumption-probability}. Consumer $c_j$'s consumption probability in $t$ is computed according to
\begin{equation}
\label{eq:consumption-probability}
p_{jt}^{\text{con}} =
\begin{cases}
\min\left( c_{jt}^{\text{trust}} + a_t \cdot \Delta, 1  \right) & \text{if } {(c_{jt}^{\text{trust}} + c_{jt}^{\text{soc}})}/{2} > c_{jt}^{\text{trust}} + a_t \cdot \Delta  \\
\max\left( c_{jt}^{\text{trust}} - a_t \cdot \Delta, 0  \right) & \text{if } {(c_{jt}^{\text{trust}} + c_{jt}^{\text{soc}})}/{2} < c_{jt}^{\text{trust}} - a_t \cdot \Delta  \\
{(c_{jt}^{\text{trust}} + c_{jt}^{\text{soc}})}/{2} & \text{otherwise.}
\end{cases}
\end{equation}

\noindent \textcolor{black}{The cases included in Eq. \ref{eq:consumption-probability} make sure that the consumption probability can take values in the shaded area in Fig. \ref{fig:consumption-probability}: The first and the second case set the upper an lower limits, respectively. Whenever the mean of consumer $c_j$'s trust and the provider's social reputation exceed one of the boundaries, the consumption probability takes the exact value of the boundary. The third case captures situations in which the mean of consumer $c_j$'s consumption probability and the provider's social reputation is located in the shaded area and, therefore, takes the value of the mean.}

\paragraph{\textcolor{black}{Item choice}}
Every consumer decides whether to purchase an item or not every time they receive a recommendation. Consumers receive a list of $\mathcal{N}$ items as recommendation in every time step. To keep our model simple, we limit the consumption to one item per time step for each consumer. \textcolor{black}{Consumers can however select any of the $\mathcal{N}$ recommendations. In reality, consumer choices can be influenced by \emph{position biases}, and items higher up in a list receive more attention and may be selected more frequently. To account for this phenomenon in our model, we assume that the selection probability for an item in the list decreases with its position. Technically, we model this by assigning probabilities $p_{k}^{\text{list}}$ to the items based on their ranks. The probabilities are drawn from a Dirichlet distribution, where $k=(1,\dots,\mathcal{N})$. We use the Dirichlet distribution to create probability mass functions that represent consumption probabilities for finite sets of recommended items. This means, we use the distribution to create $\mathcal{N}$ positive probabilities that add up to one.\footnote{More technically, the probabilities fulfill the conditions $p_{k}^{\text{list}} > 0$ and $\sum_{k=1}^\mathcal{N} p_{k}^{\text{list}} = 1$.} Then, we order the probabilities in a descending order, and assign them to the items on the recommendation list. In consequence, the item ranked first (last) on the recommendation list gets consumed with the highest (lowest) probability.}

\paragraph{Discussion}
In this section, we have presented a novel approach to model consumer behavior in a realistic way. Our behavioral model is based on the assumption that the consumption probability of consumers and their trust in a service depends both on their past experiences as well as on the public reputation of the provider. Generally, the behavior model is designed in a way that the beliefs of consumers change gradually and smoothly over time. A number of parameters were furthermore introduced in the model to accommodate application-specific idiosyncrasies. We are not aware of earlier behavior models from the literature, which consider the effects of different quality levels of the recommendations on long-term consumer beliefs in a similar way.

\section{Simulation Results and Discussion}
\label{sec:results}

This section reports the outcomes of an extensive set of simulations based on the model described in the previous section.

\subsection{Simulation Settings}
\label{subsec:simluation-settings}

Let us first summarize the main input settings and outcome variables of our simulations according to our model. To investigate the long-term effects of different recommendation strategies---which is the primary goal of our research---we observe the following outcome (dependent) variables over time:

\begin{itemize}
\item  The average \emph{consumption probability} of consumers, which can be influenced both by the consumers' trust based on their own experience and the ``reputation'' of the provider on social media.
\item The \emph{profit} that results from item consumption triggered by recommendations by all agents \emph{in each step}.
\item The \emph{cumulative} profit that accumulates over time.
\end{itemize}

The primary input (independent) variable is the choice of the recommendation strategy (consumer-centric, profit-centric, balanced, consumer-biased, popularity-based)as outlined in Section~\ref{sec:simulation-model-details}.
Moreover, through our simulations, we aim to explore the effects of assuming different levels for two essential model parameters, namely \emph{(a)} the extent consumers rely on social media during trust formation and \emph{(b)} the expectations that consumers have regarding the relevance of the recommendations they receive.

\begin{table}[h!t]
\caption{Main inputs and output of the performed simulations.}
\label{tab:model-parameters}
\centering
\begin{tabular}{ll}
\toprule
Variable & Parameters\\ \midrule
\textbf{Output (dependent) variables:}  & \\
Profit per time step   &    \\
Cumulative profit  & \\
Average trust &  \\
Average consumption probability &  \\
{\textbf{Input (independent) variables:}}     &   \\
Recommendation strategies             &   \\
\mbox{~}--- Consumer-centric                    & $\omega = 1$   \\
\mbox{~}--- Profit-centric                      & $\omega = 0$   \\
\mbox{~}--- Consumer-biased                     & $\omega = 0.9$  \\
\mbox{~}--- Balanced                            & $\omega = 0.5$  \\
\mbox{~}--- Popularity-based                    &   \\
{\textbf{Further variables:}}     &     \\
Social reliance & $\Delta=\{0, 0.05, 0.1, 0.5\} $  \\
Consumer expectation quantile  &$\psi = \{0.75, 0.85, 0.95\}$ \\
\bottomrule
\end{tabular}
\end{table}

Overall, given these input and control variables, we conducted 60 simulations to explore the effects on the performance of five recommendation strategies, four levels of reliance on social media, and three levels of consumer expectations. A summary of the main model variables and parameters for these simulations is provided in Tab.~\ref{tab:model-parameters}.
Further, the simulation model described in Sec.~\ref{sec:model} considers several additional parameters and thresholds, which we kept constant in our simulations. The specific values or distributions that we use in the simulations are reported in Tab.~\ref{tab:static-model-parameters}. Recall, we initialize the model based on the MovieLens dataset; consequently, there are $610$ consumers (agents) and $9,742$ items. We observe the dependent variables throughout $1,000$ time steps. We run each simulation three times to minimize random effects, compute the mean of each model's outcome, and show the confidence interval with $\alpha = 0.05$ in all figures.

\begin{table}[h!t]
\centering
\caption{Specific (static) simulation settings.}
\label{tab:static-model-parameters}
\begin{tabular}{@{}p{9.5cm}lp{2.1cm}@{}}
\toprule
Parameter & Symbol & Value/Distr. \\
\midrule
\textbf{Provider:} &  &  \\
Noise in predicting consumers' utilities & $\epsilon$ & ${N}(0,0.3)$ \\
Periods between updates of expectations & $T_u$ & 100 \\
Number of items recommended to the consumers per time step & $\mathcal{N}$ & 10 \\
\textbf{Consumers:} &  &  \\
Number of consumers & $J$ & 610 \\
Probability to observe social media information & $p^\text{soc}$ & 0.15 \\
Probability to provide feedback to the provider & $p^\text{feed}$ & 0.10 \\
Social media volume threshold & $\nu$ & $5,000$ \\
\textbf{Items:} &  &  \\
Number of items & $\textcolor{black}{M}$ & $9,742$ \\
Profit per item & $\rho_i$ & $N(2.5,1), \newline 0 \leq \rho_i \leq 5$\\
\midrule
\textbf{Simulation process settings:} &  &  \\
Simulation time steps & $T$ & $1,000$ \\
Replications & $N$ & 3 \\ \bottomrule
\end{tabular}
\end{table}

\subsection{Results: Effects of Recommendation Strategies without Social Media}
\label{subsec:results-no-social-media}
\begin{figure}
     \centering
     \begin{subfigure}{0.3\textwidth}
         \centering
         \includegraphics [width=\textwidth]{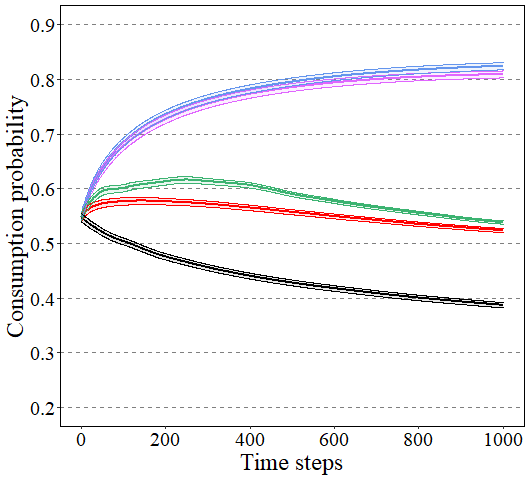}
         \caption{Consumption probability}
         \label{fig:no-social-media-consumption}
     \end{subfigure}
     \hspace{0.1em}
     \begin{subfigure}{0.3\textwidth}
         \centering
         \includegraphics[width=\textwidth]{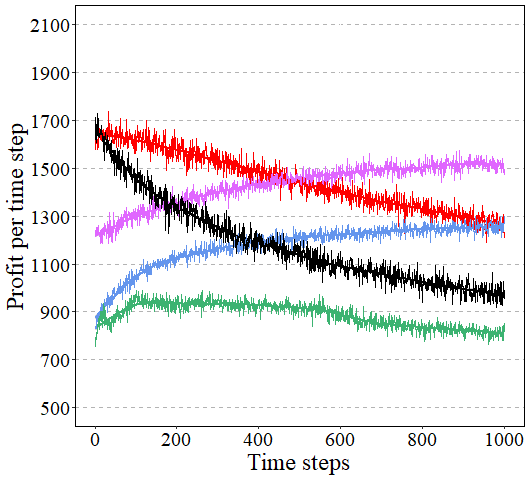}
         \caption{Profit per time step}
         \label{fig:no-social-media-profit-per-step}
     \end{subfigure}
     \hspace{0.1em}
     \begin{subfigure}{0.3\textwidth}
         \centering
         \includegraphics[width=\textwidth]{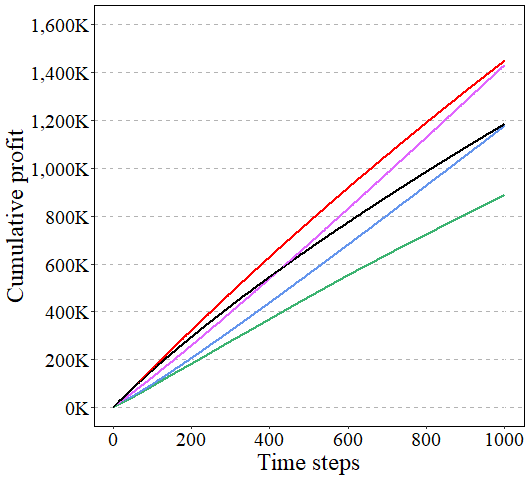}
         \caption{Cumulative profit}
         \label{fig:no-social-media-cumulative-profit}
     \end{subfigure}
    \par\bigskip
      \begin{subfigure}{0.8\textwidth}
         \centering
         \includegraphics[width=0.8\textwidth]{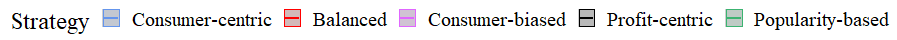}
     \end{subfigure}
        \caption{Effects without considering social media (mean with $\pm{95}$\,\% CIs).}
        \label{fig:no-social-media}
\end{figure}

This section first turns the attention to a model that does not consider information sharing on social media. In these cases, the consumers' trust in the provider---and thus their consumption probability---solely depends on their own experiences with the recommendations. Furthermore, we consider a consumer expectation quantile of $\psi$ of 0.75. Figure~\ref{fig:no-social-media} includes the average consumption probability (Fig.~\ref{fig:no-social-media-consumption}), profit per time step (Fig.~\ref{fig:no-social-media-profit-per-step}), and cumulative profit (Fig.~\ref{fig:no-social-media-cumulative-profit}). The results for further expectation quantiles are included in the \textcolor{black}{Appendix}. 

Regarding the \emph{consumption probability} (which is equal to trust in this scenario) (Fig.~\ref{fig:no-social-media-consumption}), we find that the consumer-centric and consumer-biased strategies perform best. They converge to levels between 0.8 and 0.85, and the probability starts to stabilize at these levels at the end of the observation period. Please note that we do not expect the consumption probability to reach the maximum value of 1 because of various stochastic model components and the ``natural'' noise in predicting consumer utilities. For the popularity-based strategy, we observe that the probability initially increases to a level of around 0.62. However, after only about 150 periods, it continuously decreases. Thus, since consumption in the popularity-based strategy is at intermediate levels, recommending popular items without considering consumer preferences does not strongly ``hurt'' trust and the consumption probability. Still, non-personalized recommendations do not help develop it over time either, as consumers often receive recommendations that do not meet their expectations. We observe a similar pattern for the (profit-utility) \emph{balanced} strategy.

We observe the lowest consumption probability over time in the case of the profit-based strategy. Recall, considering item profitability only results in random recommendations that do not consider the consumers' preferences. Consequently, the recommendations are much more often not useful for consumers, and the consumption almost immediately drops. Again, due to the architecture of the proposed model, we do not expect it to drop to zero soon. Instead, the consumption will stabilize at relatively low levels. If consumption is low, consumers barely make new experiences with the provider. Thus, the consumption probability decreases only slowly in later periods, and increasing it once it is at low levels may be challenging.

We finally see that the popularity-based strategy performs significantly better than the profit-based strategy, even though both strategies do not consider the consumers' preferences. We explain this pattern by the randomly assigned profit values, because there will always be profitable and relevant items for some users.

When we look at the sum of \emph{profit in each time step} (Fig.~\ref{fig:no-social-media-profit-per-step}), we find that the purely profit-oriented and balanced strategies initially lead to the highest profit.\footnote{{The jagged lines in Fig.~\ref{fig:no-social-media-profit-per-step} are the result of the random factors in the simulation.}} 
However, the profit per time step drops when consumers build up negative experiences with the provider because they consume items from their recommendation lists that are only of little utility for them. In consequence, trust and consumption probability decrease, and, as a result, they consume items less often. This drop is more pronounced for the purely profit-oriented strategy than for the balanced strategy. In contrast, for the consumer-oriented strategies, the profit per time step constantly increases. The increase in profit is driven by the growing trust in the recommendation provider. After some time, the profit per time step finally exceeds the profit generated with strategies that focus more strongly on profit. The profit per time step for the non-personalized popularity-based strategy is the lowest throughout the observation period. This pattern is driven by the continuous drop in trust.

The \textit{cumulative profit} is plotted in Fig.~\ref{fig:no-social-media-cumulative-profit}. Given our simulation settings, we see that more profit-oriented approaches appear to be more favorable in the short term, which is expected from Fig.~\ref{fig:no-social-media-profit-per-step}. However, \textcolor{black}{in the long run, the profit-oriented strategy performs significantly worse compared to the consumer-oriented strategy. In particular, when comparing the cumulative profit of the consumer-oriented and the profit-oriented strategies after $1,000$ time steps, focusing exclusively on profit when computing recommendations leads to a relative loss of approximately $26\,\%$}.
Over time, the highest overall profit is accumulated when both consumer utility and item profitability are considered. At the end of the simulation, the \textcolor{black}{cumulative profits} in the cases of the consumer-biased and the balanced strategies \textcolor{black}{exceed the profit of the consumer-oriented strategy by approximately $20\,\%$}. \textcolor{black}{Interestingly}, when looking at the development of consumer trust and the consumption probability (Fig.~\ref{fig:no-social-media-consumption}), we find that the corresponding levels with the balanced strategy are significantly lower than with the consumer-biased strategy. However, re-building consumer trust can be challenging, which suggests that both the cumulative profit and the consumption frequencies should be monitored. The latter would otherwise decrease over time without being noticed.

\subsection{Results: Effects of Recommendation Strategies considering Social Media}
\label{subsec:results-with-social-media}
\begin{figure}
     \centering
     \begin{subfigure}{0.45\textwidth}
         \centering
         \includegraphics [width=\textwidth]{results/social_reliance0/consumption}
         \caption{$\Delta = 0$}
         \label{fig:with-social-reliance0-consumption}
     \end{subfigure}
     \hspace{0.1em}
     \begin{subfigure}{0.45\textwidth}
         \centering
         \includegraphics[width=\textwidth]{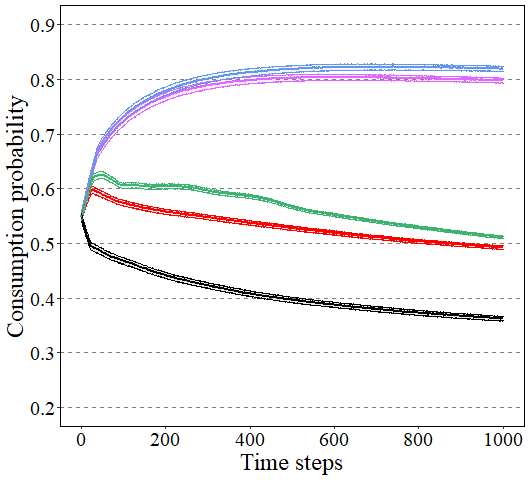}
         \caption{$\Delta = 0.05$}
         \label{fig:with-social-reliance0.05-consumption}
     \end{subfigure}

 \par\bigskip
     \begin{subfigure}{0.45\textwidth}
         \centering
         \includegraphics[width=\textwidth]{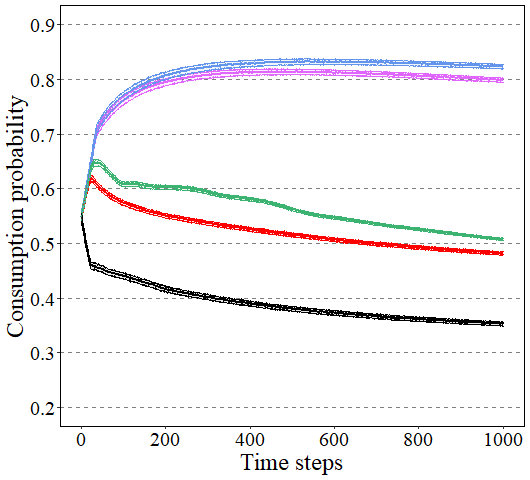}
         \caption{$\Delta = 0.1$}
         \label{fig:with-social-reliance0.1-consumption}
     \end{subfigure}
 \hspace{0.1em}
\begin{subfigure}{0.45\textwidth}
         \centering
         \includegraphics[width=\textwidth]{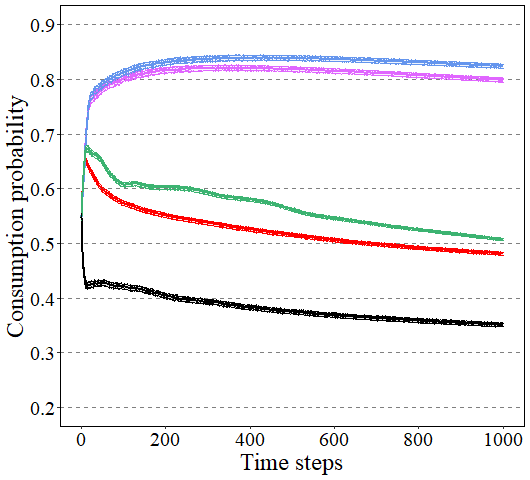}
         \caption{$\Delta = 0.5$}
         \label{fig:with-social-reliance0.5-consumption}
     \end{subfigure}
     \par\bigskip
      \begin{subfigure}{0.8\textwidth}
         \centering
         \includegraphics[width=0.8\textwidth]{results/legend}
     \end{subfigure}
        \caption{Effects on consumers' consumption probabilities with different levels of reliance on social media.}
     \label{fig:consumption-with-social-reliance}

\end{figure}

We now focus on the effects of recommendation strategies when social media comes into play, i.e., when the individual consumption probabilities are not only based on the agents' personal experiences but are also positively or negatively affected by what other agents share on social media regarding their experiences with the recommendations. Figure~\ref{fig:consumption-with-social-reliance} shows the \emph{consumption probabilities} for different levels of influence by social media (from 0 to 0.5). Note that the setting of zero in Fig. \ref{fig:with-social-reliance0-consumption} corresponds to not considering social media at all, and the results, therefore, correspond to those from the previous section.

Overall, we observe that social media has a \emph{reinforcing} effect. For the more consumer-oriented strategies, we see that the consumption probability grows faster. Recall, these strategies provide consumers with recommendations that better meet their expectations. Consequently, the consumers are satisfied and share rather positive than negative information on social media. Once available on social media, this information is considered by other consumers when making consumption decisions (and increases their probability of consumption). Of course, information shared on social media might also unfold more negative effects: for
the other strategies, we observe that at the end of the simulation, i.e., after $1,000$ time steps, the average consumption probabilities are only slightly lower than in situations when consumers do not consider social media postings. However, the consumption probability drops almost immediately in early time steps.

\begin{figure}
     \centering
     \begin{subfigure}{0.45\textwidth}
         \centering
         \includegraphics [width=\textwidth]{results/social_reliance0/profit-per-step}
         \caption{$\Delta = 0$}
         \label{fig:with-social-reliance0-profit-per-step}
     \end{subfigure}
     \hspace{0.1em}
     \begin{subfigure}{0.45\textwidth}
         \centering
         \includegraphics[width=\textwidth]{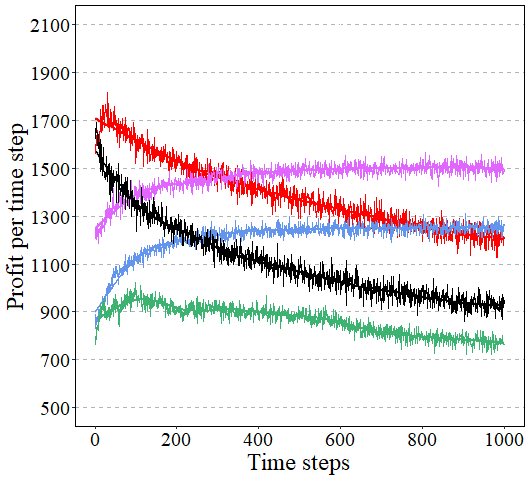}
         \caption{$\Delta = 0.05$}
         \label{fig:with-social-reliance0.05-profit-per-step}
     \end{subfigure}

 \par\bigskip

     \begin{subfigure}{0.45\textwidth}
         \centering
         \includegraphics[width=\textwidth]{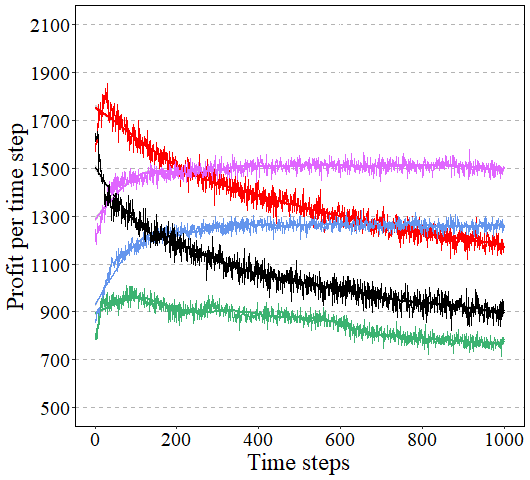}
         \caption{$\Delta = 0.1$}
         \label{fig:with-social-reliance0.1-profit-per-step}
     \end{subfigure}
 \hspace{0.1em}
\begin{subfigure}{0.45\textwidth}
         \centering
         \includegraphics[width=\textwidth]{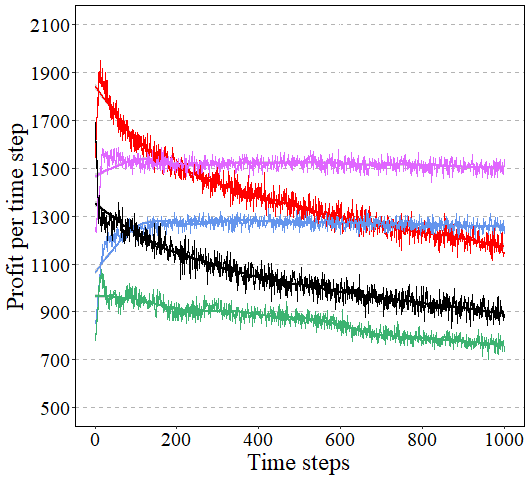}
         \caption{$\Delta = 0.5$}
         \label{fig:with-social-reliance0.5-profit-per-step}
     \end{subfigure}
     \par\bigskip
      \begin{subfigure}{0.8\textwidth}
         \centering
         \includegraphics[width=0.8\textwidth]{results/legend}
     \end{subfigure}
        \caption{Effects on the profit per time step with different levels of reliance on social media.}
       \label{fig:profit-per-step-with-social-reliance}

\end{figure}

\begin{figure}
     \centering
     \begin{subfigure}{0.45\textwidth}
         \centering
         \includegraphics [width=\textwidth]{results/social_reliance0/cumulative-profit}
         \caption{$\Delta = 0$}
         \label{fig:with-social-reliance0-cumulative-profit}
     \end{subfigure}
     \hspace{0.1em}
     \begin{subfigure}{0.45\textwidth}
         \centering
         \includegraphics[width=\textwidth]{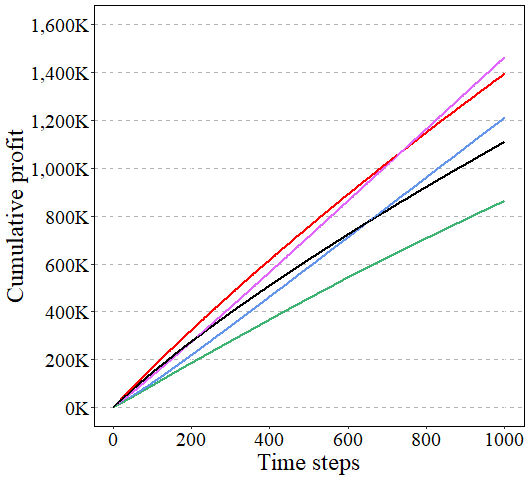}
         \caption{$\Delta = 0.05$}
         \label{fig:with-social-reliance0.05-cumulative-profit}
     \end{subfigure}
 \par\bigskip
     \begin{subfigure}{0.45\textwidth}
         \centering
         \includegraphics[width=\textwidth]{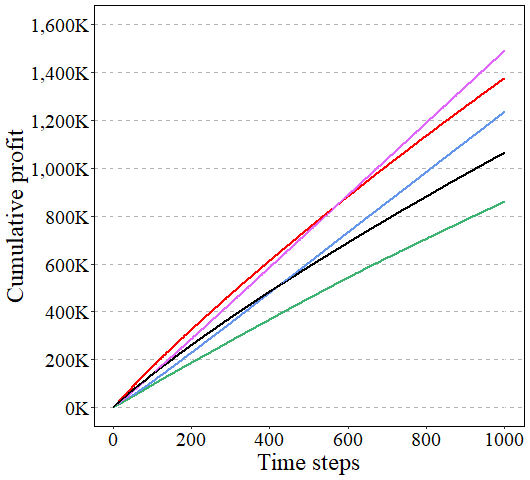}
         \caption{$\Delta = 0.1$}
         \label{fig:with-social-reliance0.1-cumulative-profit}
     \end{subfigure}
 \hspace{0.1em}
\begin{subfigure}{0.45\textwidth}
         \centering
         \includegraphics[width=\textwidth]{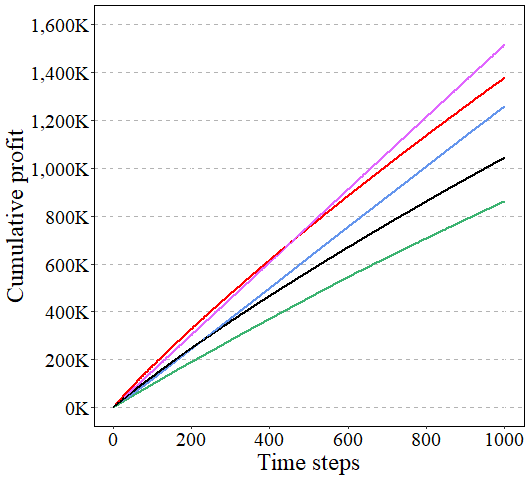}
         \caption{$\Delta = 0.5$}
         \label{fig:with-social-reliance0.5-cumulative-profit}
     \end{subfigure}
     \par\bigskip
      \begin{subfigure}{0.8\textwidth}
         \centering
         \includegraphics[width=0.8\textwidth]{results/legend}
     \end{subfigure}
        \caption{Effects on the cumulative profit with different levels of reliance on social media.}
    \label{fig:cumulative-profit-with-social-reliance}
\end{figure}

The profit per time step and the cumulative profit are plotted in Figs.~\ref{fig:profit-per-step-with-social-reliance} and \ref{fig:cumulative-profit-with-social-reliance}, respectively. Looking at the profits per time step in Fig.~\ref{fig:profit-per-step-with-social-reliance}, we observe that information sharing via social media unfolds a reinforcing effect in the early phases of the simulation. Recall, depending on the recommendation strategy, the consumption probabilities increase or decrease more quickly as social media comes into play. The profits per time step follow this pattern. Finally, concerning the cumulative profit in Fig.~\ref{fig:cumulative-profit-with-social-reliance}, the results indicate that the consumer-biased strategy leads to the highest cumulative profit at the end of the simulation if consumers consider the information on social media when making decisions. \textcolor{black}{In particular, the cumulative profit of the consumer-biased strategy exceeds that of the consumer-oriented strategy by approximately $20\,\%$ for all levels of reliance on social media.} At the same time, this strategy preserves high levels of trust (and thus, consumption probabilities), as shown in Fig.~\ref{fig:consumption-with-social-reliance}. \textcolor{black}{The balanced strategy also leads to a higher cumulative profit compared to the consumer-oriented strategy for all levels of reliance on social media. Please note that the difference in cumulative profit \textit{decreases} as the consumers' reliance on social media increases. For example, for $\Delta = 0.05$, the difference in cumulative profit amounts to approximately $15\,\%$. If consumers heavily rely on social media when making decisions ($\Delta=0.5$), the difference in cumulative profit only amounts to approximately $9\,\%$. Finally, the popularity-based and profit-oriented strategies lead to more severe losses if consumers more heavily rely on the provider's social reputation when making decisions. For example, comparing the cumulative profits in the cases of the popularity-based and the consumer-oriented strategies reveals that the difference between the profits increases from $-28\,\%$ to $-31\,\%$ when the consumers' reliance on social media increases from $\Delta=0.05$ to $\Delta=0.5$. For the comparison of the profit-oriented and the consumer-based strategies, this effect is even more pronounced. In this case, the difference in cumulative profit is $-8\,\%$ when $\Delta=0.05$ and $-17\,\%$ when $\Delta=0.5$.}

The simulation of the long-term effects of different provider strategies presented in this paper shows that \emph{incorporating profitability considerations in the recommendation process may help to create sustained business success}, and, most importantly, this may be achieved without significantly compromising the consumers' satisfaction with the recommendation service in the long run.\footnote{Recall, we model the consumption probability to be mainly dependent on trust plus some external effect, and the latter is driven by information shared on social media. Hence, the consumption probability is a good proxy for how satisfied the consumers are with the provided recommendation service.} \textcolor{black}{However, we also show that focusing on short-term profitability might unfold adverse effects in the long run.} Our approach is among the first to study the longitudinal effects of recommender systems that consider both the perspectives of the provider and the consumer. Previous attempts to design profit-maximizing recommender systems often only implicitly consider both perspectives. For example, \cite{pathak2010} aim to study the interrelations between recommendations and sales; however, the effect on sales is only considered indirect (i.e., what gets recommended more might be bought more). \cite{das2009maximizing} follow a mathematical approach to profit-maximizing recommender systems, and, at a conceptual level, they argue that recommending similar items with higher profit margins can increase profit.

More closely related to our research objective, \cite{hosanagar2008}, \cite{chen2008}, and \cite{azaria2013} argue that (personalized) recommendations that take into account profitability aspects can be beneficial for providers. In addition to their work, our analysis provides additional insights into the longitudinal effects. In particular, we emphasize that profit-aware recommendation strategies must be designed with care and closely monitored longitudinally. Such strategies, if tuned too much towards profitability, may \emph{(a)} lead to a lowered quality perception of the recommendations by the consumers; and they may \emph{(b)} negatively influence the consumers' attitudes towards the recommendation service, and ultimately the provider's profitability, in the long run. According to our simulations, the most promising choice for providers in the long-term is to primarily focus on consumers' value but to consider profitability as a secondary factor. The second insight of our simulations is that public posts on consumers on social media about their quality perceptions may lead to reinforcing effects. The observed effect is consistent with the concept of bottom-up marketing in \cite{karpinski2005} and \cite{hanna2011}, i.e., user-generated content is created, and consumers trust the opinions of their peers. Consequently, a reinforcing effect emerges. This effect may happen in both directions: many positive social media posts help increase consumption, whereas negative posts may lead to a rapid drop in consumption, thereby affecting profitability. Thus, the observed social media effects also emphasize the first insight's importance.

Overall, we see our work as an essential step towards a better understanding of the longitudinal effects of recommender systems for consumers---which is the traditional focus of research---and for different stakeholders. 
Our proposed \emph{\textcolor{black}{agent-based}} model allows us to model the behavior of consumers on an individual (micro-)level, where consumers adapt their behavior over time, based, e.g., on their expectations towards the quality of the recommendations, their experienced quality over time, and the reputation of the provider in a community. The collective behavior of these individuals then lets us study emerging and longitudinal phenomena on the macro-level.

\section{Conclusion and Future Work}
\label{sec:conclusion}

Most existing research in the field of recommender systems aims to design algorithms that maximize the relevance of the presented recommended items for consumers in the short term. Only in recent years the multi-sided nature of the problem has moved in the research focus, where researchers strive to design recommendation strategies that can consider the value creation perspective of multiple stakeholders. Moreover, it has become apparent that optimizing only for the short-term might lead to undesired effects in the long run. With this work, we propose a simulation model that addresses these problems and allows us to study emerging longitudinal effects of different recommendation strategies cost-efficiently. Furthermore, we share the simulation framework that we used as an open source-solution to be freely used by other researchers.

A variety of extensions to our model are possible. For example, we currently assume a static item catalog, as was done in \cite{consumptionPerformance2020}. In reality, however, there are many domains where new items constantly come in, e.g., new movies or shop items, and where others become out-of-stock or are generally no longer available at some stage. Such dynamics may have a relevant impact on specific recommendation strategies, e.g., when the recency or trendiness of an item impacts the quality perception of users or when high-profit items cannot be recommended at certain times.

Looking at the dynamics of information exchange in our model, we currently consider feedback from consumers to providers and consumers to social media. Other forms of information exchange, e.g., directly between the consumers, are so far left for future work. Moreover, the model investigated in our present work only considers the dynamics that may evolve between one provider (and its recommendation strategy) and a set of consumers. It would be interesting to explore the dynamics in markets when there is more than one provider in future work. In such a case, consumers' behaviors may depend not only on the experiences with the current provider but also on quality expectations for an alternative provider, e.g., based again on social media. Also, in the current model, we assume that customers cannot drop out of the recommendation service, but only their consumption probability decreases. Models with multiple providers and customers that can drop out of (one or more) of the provided recommendation services may also allow analyzing phenomena like provider switches or the reliance on multiple providers in parallel, e.g., for different shop categories.

Independent of the market model, i.e., if there are one or more providers, we may explore alternative recommendation strategies in the future. Currently, our providers are static because they do not change their strategy during the simulation. In reality, however, providers might closely monitor the effects of the recommendations on consumer behavior and adapt their strategies over time or employ multiple strategies simultaneously. For example, when a significant drop in consumption is observed after implementing a more profit-oriented strategy, providers might revise the strategy to re-built trust, at least for the affected consumers.

Considering how we model profitability, we currently rely on synthetic data and assign random profitability values to the available items. In some domains, however, there might be a correlation between item profitability and other factors. More frequently bought (popular) items might in some domains have a smaller profit margin than others. In the movie domain, service providers might for example have to be pay more royalties for certain types of items, e.g., more recent releases, which at the same time may be of higher value for consumers. In our future works, we plan to investigate if alternative recommendation strategies are needed to consider such correlations in an optimal way.

So far, our model does not consider end consumer prices, and we do not make the assumption that items with a higher profit margin have higher prices in general. In our simulations in the movie recommendation domain, which is the focus of our study, we currently assume a flat-rate business model by the service provider as is common for modern streaming providers.
However, in other domains, consumer prices, which may or not be correlated with provider profitability, can be important factors in the consumer decision process. The consideration of such effects however goes beyond the scope of our current study and is left for future work.

In that context, a limitation of our simulations so far is that is currently limited to one particular application scenario (movie recommendation). While we initialized the model based on consumer preferences with real-world data, we had to base the modeling of profitability aspects on synthetic data. Likewise, since we are not aware of any dataset that contains both consumer preference data and consumer activity on social media, we modelled the social feedback behavior of agents based on existing theories and intuitive assumptions. A further validation of our model in other domains and with other types of empirical data is therefore still required. In the best case, such data includes information about consumer behavior, consumer satisfaction, their behavior on social media, their trust towards service providers, or even interactions between consumers.

Despite these limitations, we believe that our work may represent an important step forward towards the realistic simulation of complex business environments, and we hope that our work fosters more research in this crucial area in the future.

\newpage

\section*{Conflict of interest}
\noindent The authors declare that they have no conflict of interest.

\section*{Code availability}
\noindent The source code is available \href{https://github.com/nadaa/rec-strategies-abm/}{here}.

\section*{Data availability}
\noindent Simulation data are available \href{https://seafile.aau.at/f/1e84eec700244ac1871d/}{here}.

\bibliography{bibfile}

\begin{thebibliography}{67}
\expandafter\ifx\csname natexlab\endcsname\relax\def\natexlab#1{#1}\fi
\providecommand{\url}[1]{\texttt{#1}}
\providecommand{\href}[2]{#2}
\providecommand{\path}[1]{#1}
\providecommand{\DOIprefix}{doi:}
\providecommand{\ArXivprefix}{arXiv:}
\providecommand{\URLprefix}{URL: }
\providecommand{\Pubmedprefix}{pmid:}
\providecommand{\doi}[1]{\href{http://dx.doi.org/#1}{\path{#1}}}
\providecommand{\Pubmed}[1]{\href{pmid:#1}{\path{#1}}}
\providecommand{\bibinfo}[2]{#2}
\ifx\xfnm\relax \def\xfnm[#1]{\unskip,\space#1}\fi
\bibitem[{Abdollahpouri et~al.(2020)Abdollahpouri, Adomavicius, Burke, Guy,
  Jannach, Kamishima, Krasnodebski \& Pizzato}]{abdollahpouri2020}
\bibinfo{author}{Abdollahpouri, H.}, \bibinfo{author}{Adomavicius, G.},
  \bibinfo{author}{Burke, R.}, \bibinfo{author}{Guy, I.},
  \bibinfo{author}{Jannach, D.}, \bibinfo{author}{Kamishima, T.},
  \bibinfo{author}{Krasnodebski, J.}, \& \bibinfo{author}{Pizzato, L.}
  (\bibinfo{year}{2020}).
\newblock \bibinfo{title}{Multistakeholder recommendation: Survey and research
  directions}.
\newblock {\it \bibinfo{journal}{User Modeling and User-Adapted
  Interaction}\/},  {\it \bibinfo{volume}{30}\/}, \bibinfo{pages}{127--158}.
\bibitem[{Adomavicius et~al.(2013)Adomavicius, Gupta, Ketter \&
  Zhang}]{adomavicius2013understanding}
\bibinfo{author}{Adomavicius, G.}, \bibinfo{author}{Gupta, A.},
  \bibinfo{author}{Ketter, W.}, \& \bibinfo{author}{Zhang, J.}
  (\bibinfo{year}{2013}).
\newblock \bibinfo{title}{Understanding longitudinal dynamics of recommender
  systems performance: An agent-based modeling approach}.
\newblock In {\it \bibinfo{booktitle}{Proceedings of the 23rd Workshop on
  Information Technology and Systems: Leveraging Big Data Analytics for
  Societal Benefits}\/}.
\bibitem[{Adomavicius et~al.(2021)Adomavicius, Jannach, Leitner \&
  Zhang}]{adomavicius2021}
\bibinfo{author}{Adomavicius, G.}, \bibinfo{author}{Jannach, D.},
  \bibinfo{author}{Leitner, S.}, \& \bibinfo{author}{Zhang, J.}
  (\bibinfo{year}{2021}).
\newblock \bibinfo{title}{Understanding longitudinal dynamics of recommender
  systems with agent-based modeling and simulation}.
\newblock \bibinfo{howpublished}{arXiv preprint arXiv:2108.11068}.
\newblock \URLprefix \url{https://arxiv.org/abs/2108.11068}.
\bibitem[{Adomavicius \& Tuzhilin(2005)}]{adomavicius2005toward}
\bibinfo{author}{Adomavicius, G.}, \& \bibinfo{author}{Tuzhilin, A.}
  (\bibinfo{year}{2005}).
\newblock \bibinfo{title}{Toward the next generation of recommender systems: A
  survey of the state-of-the-art and possible extensions}.
\newblock {\it \bibinfo{journal}{IEEE Transactions on Knowledge and Data
  Engineering}\/},  {\it \bibinfo{volume}{17}\/}, \bibinfo{pages}{734--749}.
\bibitem[{Azaria et~al.(2013)Azaria, Hassidim, Kraus, Eshkol, Weintraub \&
  Netanely}]{azaria2013}
\bibinfo{author}{Azaria, A.}, \bibinfo{author}{Hassidim, A.},
  \bibinfo{author}{Kraus, S.}, \bibinfo{author}{Eshkol, A.},
  \bibinfo{author}{Weintraub, O.}, \& \bibinfo{author}{Netanely, I.}
  (\bibinfo{year}{2013}).
\newblock \bibinfo{title}{Movie recommender system for profit maximization}.
\newblock In {\it \bibinfo{booktitle}{Proceedings of the 7th ACM Conference on
  Recommender Systems}\/} (pp. \bibinfo{pages}{121--128}).
\bibitem[{Bennett et~al.(2007)Bennett, Lanning et~al.}]{bennett2007netflix}
\bibinfo{author}{Bennett, J.}, \bibinfo{author}{Lanning, S.} et~al.
  (\bibinfo{year}{2007}).
\newblock \bibinfo{title}{{The Netflix Prize}}.
\newblock In {\it \bibinfo{booktitle}{{Proceedings of KDD Cup and
  Workshop}}\/}.
\bibitem[{Bhattacherjee(2001)}]{bhattacherjee2001understanding}
\bibinfo{author}{Bhattacherjee, A.} (\bibinfo{year}{2001}).
\newblock \bibinfo{title}{Understanding information systems continuance: An
  expectation-confirmation model}.
\newblock {\it \bibinfo{journal}{MIS Quarterly}\/},  {\it
  \bibinfo{volume}{25}\/}, \bibinfo{pages}{351--370}.
\bibitem[{Bountouridis et~al.(2019)Bountouridis, Harambam, Makhortykh, Marrero,
  Tintarev \& Hauff}]{bountouridis2019siren}
\bibinfo{author}{Bountouridis, D.}, \bibinfo{author}{Harambam, J.},
  \bibinfo{author}{Makhortykh, M.}, \bibinfo{author}{Marrero, M.},
  \bibinfo{author}{Tintarev, N.}, \& \bibinfo{author}{Hauff, C.}
  (\bibinfo{year}{2019}).
\newblock \bibinfo{title}{{SIREN}: A simulation framework for understanding the
  effects of recommender systems in online news environments}.
\newblock In {\it \bibinfo{booktitle}{Proceedings of the Conference on
  Fairness, Accountability, and Transparency}\/} (pp.
  \bibinfo{pages}{150--159}).
\bibitem[{Chen et~al.(2008)Chen, Hsu, Chen \& Hsu}]{chen2008}
\bibinfo{author}{Chen, L.-S.}, \bibinfo{author}{Hsu, F.-H.},
  \bibinfo{author}{Chen, M.-C.}, \& \bibinfo{author}{Hsu, Y.-C.}
  (\bibinfo{year}{2008}).
\newblock \bibinfo{title}{Developing recommender systems with the consideration
  of product profitability for sellers}.
\newblock {\it \bibinfo{journal}{Information Sciences}\/},  {\it
  \bibinfo{volume}{178}\/}, \bibinfo{pages}{1032--1048}.
\bibitem[{Chen et~al.(2021)Chen, Chen \& Jin}]{chen2021}
\bibinfo{author}{Chen, T.}, \bibinfo{author}{Chen, Z.}, \&
  \bibinfo{author}{Jin, X.} (\bibinfo{year}{2021}).
\newblock \bibinfo{title}{A multiple information model incorporating limited
  attention and information environment}.
\newblock {\it \bibinfo{journal}{PLOS ONE}\/},  {\it \bibinfo{volume}{16}\/},
  \bibinfo{pages}{e0257844}.
\bibitem[{Das et~al.(2009)Das, Mathieu \& Ricketts}]{das2009maximizing}
\bibinfo{author}{Das, A.}, \bibinfo{author}{Mathieu, C.}, \&
  \bibinfo{author}{Ricketts, D.} (\bibinfo{year}{2009}).
\newblock \bibinfo{title}{Maximizing profit using recommender systems}.
\newblock \bibinfo{howpublished}{arXiv preprint arXiv:0908.3633}.
\newblock \URLprefix \url{https://arxiv.org/abs/0908.3633v1}.
\bibitem[{Dinh \& Parulian(2020)}]{dinh2020}
\bibinfo{author}{Dinh, L.}, \& \bibinfo{author}{Parulian, N.}
  (\bibinfo{year}{2020}).
\newblock \bibinfo{title}{{COVID-19 pandemic and information diffusion analysis
  on Twitter}}.
\newblock {\it \bibinfo{journal}{Proceedings of the Association for Information
  Science and Technology}\/},  {\it \bibinfo{volume}{57}\/},
  \bibinfo{pages}{e252}.
\bibitem[{Donkers \& Ziegler(2021)}]{donkers2021dual}
\bibinfo{author}{Donkers, T.}, \& \bibinfo{author}{Ziegler, J.}
  (\bibinfo{year}{2021}).
\newblock \bibinfo{title}{The dual echo chamber: Modeling social media
  polarization for interventional recommending}.
\newblock In {\it \bibinfo{booktitle}{Fifteenth ACM Conference on Recommender
  Systems}\/} (pp. \bibinfo{pages}{12--22}).
\bibitem[{Ekstrand et~al.(2014)Ekstrand, Harper, Willemsen \&
  Konstan}]{Ekstrand:2014:UPD:2645710.2645737}
\bibinfo{author}{Ekstrand, M.~D.}, \bibinfo{author}{Harper, F.~M.},
  \bibinfo{author}{Willemsen, M.~C.}, \& \bibinfo{author}{Konstan, J.~A.}
  (\bibinfo{year}{2014}).
\newblock \bibinfo{title}{User perception of differences in recommender
  algorithms}.
\newblock In {\it \bibinfo{booktitle}{Proceedings of the 8th ACM Conference on
  Recommender Systems}\/} RecSys '14 (pp. \bibinfo{pages}{161--168}).
\bibitem[{Epstein(2006)}]{Epstein2006}
\bibinfo{author}{Epstein, J.~M.} (\bibinfo{year}{2006}).
\newblock {\it \bibinfo{title}{{Generative social science: Studies in
  agent-based computational modeling}}\/} volume~\bibinfo{volume}{13}.
\newblock \bibinfo{address}{Princeton, NJ}: \bibinfo{publisher}{Princeton
  University Press}.
\bibitem[{Ethiraj \& Levinthal(2009)}]{ethiraj2009}
\bibinfo{author}{Ethiraj, S.~K.}, \& \bibinfo{author}{Levinthal, D.}
  (\bibinfo{year}{2009}).
\newblock \bibinfo{title}{Hoping for {A} to {Z} while rewarding only {A}:
  Complex organizations and multiple goals}.
\newblock {\it \bibinfo{journal}{Organization Science}\/},  {\it
  \bibinfo{volume}{20}\/}, \bibinfo{pages}{4--21}.
\bibitem[{Ferraro et~al.(2020)Ferraro, Jannach \&
  Serra}]{ferrarojannachserra2020recsys}
\bibinfo{author}{Ferraro, A.}, \bibinfo{author}{Jannach, D.}, \&
  \bibinfo{author}{Serra, X.} (\bibinfo{year}{2020}).
\newblock \bibinfo{title}{Exploring longitudinal effects of session-based
  recommendations}.
\newblock In {\it \bibinfo{booktitle}{Proceedings of the 2020 ACM Conference on
  Recommender Systems}\/} (p. \bibinfo{pages}{474–479}).
\bibitem[{Fleder \& Hosanagar(2009)}]{FlederBlockbuster2009}
\bibinfo{author}{Fleder, D.}, \& \bibinfo{author}{Hosanagar, K.}
  (\bibinfo{year}{2009}).
\newblock \bibinfo{title}{Blockbuster culture's next rise or fall: The impact
  of recommender systems on sales diversity}.
\newblock {\it \bibinfo{journal}{Management Science}\/},  {\it
  \bibinfo{volume}{55}\/}, \bibinfo{pages}{697--712}.
\bibitem[{Gilbert \& Terna(2000)}]{gilbert2000}
\bibinfo{author}{Gilbert, N.}, \& \bibinfo{author}{Terna, P.}
  (\bibinfo{year}{2000}).
\newblock \bibinfo{title}{How to build and use agent-based models in social
  science}.
\newblock {\it \bibinfo{journal}{Mind \& Society}\/},  {\it
  \bibinfo{volume}{1}\/}, \bibinfo{pages}{57--72}.
\bibitem[{Graham et~al.(1989)Graham, Knuth \& Patashnik}]{Graham1989}
\bibinfo{author}{Graham, R.~L.}, \bibinfo{author}{Knuth, D.~E.}, \&
  \bibinfo{author}{Patashnik, O.} (\bibinfo{year}{1989}).
\newblock {\it \bibinfo{title}{Concrete mathematics. A foundation for computer
  science}\/}.
\newblock \bibinfo{address}{Reading, MA}: \bibinfo{publisher}{Addison-Wesley}.
\bibitem[{Greco et~al.(2017)Greco, Suglia, Basile \&
  Semeraro}]{Greco2017Converse}
\bibinfo{author}{Greco, C.}, \bibinfo{author}{Suglia, A.},
  \bibinfo{author}{Basile, P.}, \& \bibinfo{author}{Semeraro, G.}
  (\bibinfo{year}{2017}).
\newblock \bibinfo{title}{Converse-et-impera: Exploiting deep learning and
  hierarchical reinforcement learning for conversational recommender systems}.
\newblock In \bibinfo{editor}{F.~Esposito}, \bibinfo{editor}{R.~Basili},
  \bibinfo{editor}{S.~Ferilli}, \& \bibinfo{editor}{F.~A.~Lisi} (Eds.), {\it
  \bibinfo{booktitle}{AI*IA 2017 Advances in Artificial Intelligence}\/} (pp.
  \bibinfo{pages}{372--386}).
\bibitem[{{GroupLens Research}(2018)}]{movielens}
[dataset] \bibinfo{author}{{GroupLens Research}} (\bibinfo{year}{2018}).
\newblock \bibinfo{title}{{MovieLens} latest dataset small}.
\newblock \URLprefix \url{https://grouplens.org/datasets/movielens/latest/}.
\bibitem[{Gurgone et~al.(2018)Gurgone, Iori \& Jafarey}]{gurgone2018}
\bibinfo{author}{Gurgone, A.}, \bibinfo{author}{Iori, G.}, \&
  \bibinfo{author}{Jafarey, S.} (\bibinfo{year}{2018}).
\newblock \bibinfo{title}{The effects of interbank networks on efficiency and
  stability in a macroeconomic agent-based model}.
\newblock {\it \bibinfo{journal}{Journal of Economic Dynamics and Control}\/},
  {\it \bibinfo{volume}{91}\/}, \bibinfo{pages}{257--288}.
\bibitem[{Gutjahr \& Pichler(2016)}]{gutjahr2016stochastic}
\bibinfo{author}{Gutjahr, W.~J.}, \& \bibinfo{author}{Pichler, A.}
  (\bibinfo{year}{2016}).
\newblock \bibinfo{title}{Stochastic multi-objective optimization: A survey on
  non-scalarizing methods}.
\newblock {\it \bibinfo{journal}{Annals of Operations Research}\/},  {\it
  \bibinfo{volume}{236}\/}, \bibinfo{pages}{475--499}.
\bibitem[{Haki et~al.(2020)Haki, Beese, Aier \& Winter}]{haki2020}
\bibinfo{author}{Haki, K.}, \bibinfo{author}{Beese, J.}, \bibinfo{author}{Aier,
  S.}, \& \bibinfo{author}{Winter, R.} (\bibinfo{year}{2020}).
\newblock \bibinfo{title}{The evolution of information systems architecture: An
  agent-based simulation model.}
\newblock {\it \bibinfo{journal}{MIS Quarterly}\/},  {\it
  \bibinfo{volume}{44}\/}, \bibinfo{pages}{155--184}.
\bibitem[{Hanna et~al.(2011)Hanna, Rohm \& Crittenden}]{hanna2011}
\bibinfo{author}{Hanna, R.}, \bibinfo{author}{Rohm, A.}, \&
  \bibinfo{author}{Crittenden, V.~L.} (\bibinfo{year}{2011}).
\newblock \bibinfo{title}{We’re all connected: The power of the social media
  ecosystem}.
\newblock {\it \bibinfo{journal}{Business Horizons}\/},  {\it
  \bibinfo{volume}{54}\/}, \bibinfo{pages}{265--273}.
\bibitem[{Harper \& Konstan(2015)}]{MovieLens2016}
\bibinfo{author}{Harper, F.~M.}, \& \bibinfo{author}{Konstan, J.~A.}
  (\bibinfo{year}{2015}).
\newblock \bibinfo{title}{{The MovieLens Datasets: History and Context}}.
\newblock {\it \bibinfo{journal}{ACM Trans. Interact. Intell. Syst.}\/},  {\it
  \bibinfo{volume}{5}\/}.
\bibitem[{Hinz \& Eckert(2010)}]{Hinz2010Impact}
\bibinfo{author}{Hinz, O.}, \& \bibinfo{author}{Eckert, J.}
  (\bibinfo{year}{2010}).
\newblock \bibinfo{title}{The impact of search and recommendation systems on
  sales in electronic commerce}.
\newblock {\it \bibinfo{journal}{Business \& Information Systems
  Engineering}\/},  {\it \bibinfo{volume}{2}\/}, \bibinfo{pages}{66--77}.
\bibitem[{Hosanagar et~al.(2008)Hosanagar, Krishnan \& Ma}]{hosanagar2008}
\bibinfo{author}{Hosanagar, K.}, \bibinfo{author}{Krishnan, R.}, \&
  \bibinfo{author}{Ma, L.} (\bibinfo{year}{2008}).
\newblock \bibinfo{title}{Recommended for you: The impact of profit incentives
  on the relevance of online recommendations}.
\newblock In {\it \bibinfo{booktitle}{Proceedings of the 2008 International
  Conference on Information Systems}\/}.
\bibitem[{Huang et~al.(2020)Huang, Oosterhuis, de~Rijke \& van
  Hoof}]{huang2020keeping}
\bibinfo{author}{Huang, J.}, \bibinfo{author}{Oosterhuis, H.},
  \bibinfo{author}{de~Rijke, M.}, \& \bibinfo{author}{van Hoof, H.}
  (\bibinfo{year}{2020}).
\newblock \bibinfo{title}{Keeping dataset biases out of the simulation: A
  debiased simulator for reinforcement learning based recommender systems}.
\newblock In {\it \bibinfo{booktitle}{Fourteenth ACM Conference on Recommender
  Systems}\/} (pp. \bibinfo{pages}{190--199}).
\bibitem[{Hug(2020)}]{Hug2020}
\bibinfo{author}{Hug, N.} (\bibinfo{year}{2020}).
\newblock \bibinfo{title}{{Surprise: A Python library for recommender
  systems}}.
\newblock {\it \bibinfo{journal}{Journal of Open Source Software}\/},  {\it
  \bibinfo{volume}{5}\/}, \bibinfo{pages}{2174}.
\bibitem[{Hutter(2011)}]{hutter2011experience}
\bibinfo{author}{Hutter, M.} (\bibinfo{year}{2011}).
\newblock \bibinfo{title}{Experience goods}.
\newblock In {\it \bibinfo{booktitle}{A Handbook of Cultural Economics, Second
  Edition}\/}.
\newblock \bibinfo{publisher}{Edward Elgar Publishing}.
\bibitem[{Ie et~al.(2019)Ie, Hsu, Mladenov, Jain, Narvekar, Wang, Wu \&
  Boutilier}]{ie2019recsim}
\bibinfo{author}{Ie, E.}, \bibinfo{author}{Hsu, C.-w.},
  \bibinfo{author}{Mladenov, M.}, \bibinfo{author}{Jain, V.},
  \bibinfo{author}{Narvekar, S.}, \bibinfo{author}{Wang, J.},
  \bibinfo{author}{Wu, R.}, \& \bibinfo{author}{Boutilier, C.}
  (\bibinfo{year}{2019}).
\newblock \bibinfo{title}{{RecSim: A Configurable Simulation Platform for
  Recommender Systems}}.
\newblock \bibinfo{howpublished}{arXiv preprint arXiv:1909.04847}.
\newblock \URLprefix \url{https://arxiv.org/abs/1909.04847}.
\bibitem[{Jager(2021)}]{jager2021}
\bibinfo{author}{Jager, W.} (\bibinfo{year}{2021}).
\newblock \bibinfo{title}{Using agent-based modelling to explore behavioural
  dynamics affecting our climate}.
\newblock {\it \bibinfo{journal}{Current Opinion in Psychology}\/},  {\it
  \bibinfo{volume}{42}\/}, \bibinfo{pages}{133--139}.
\bibitem[{Jannach \& Adomavicius(2017)}]{JannachAdomaviciusVAMS2017}
\bibinfo{author}{Jannach, D.}, \& \bibinfo{author}{Adomavicius, G.}
  (\bibinfo{year}{2017}).
\newblock \bibinfo{title}{Price and profit awareness in recommender systems}.
\newblock In {\it \bibinfo{booktitle}{Proceedings of the ACM RecSys 2017
  Workshop on Value-Aware and Multi-Stakeholder Recommendation}\/}.
\bibitem[{Jannach \& Bauer(2020)}]{jannach2021mcnamara}
\bibinfo{author}{Jannach, D.}, \& \bibinfo{author}{Bauer, C.}
  (\bibinfo{year}{2020}).
\newblock \bibinfo{title}{Escaping the {McNamara} fallacy: Towards more
  impactful recommender systems research}.
\newblock {\it \bibinfo{journal}{AI Magazine}\/},  {\it
  \bibinfo{volume}{41}\/}, \bibinfo{pages}{79--95}.
\bibitem[{Jannach \& Jugovac(2019)}]{jannachjugovactmis2019}
\bibinfo{author}{Jannach, D.}, \& \bibinfo{author}{Jugovac, M.}
  (\bibinfo{year}{2019}).
\newblock \bibinfo{title}{Measuring the business value of recommender systems}.
\newblock {\it \bibinfo{journal}{ACM Transactions on Management Information
  Systems}\/},  {\it \bibinfo{volume}{10}\/}, \bibinfo{pages}{1--23}.
\bibitem[{Jannach et~al.(2015)Jannach, Lerche, Kamehkhosh \&
  Jugovac}]{JannachLercheEtAl2015}
\bibinfo{author}{Jannach, D.}, \bibinfo{author}{Lerche, L.},
  \bibinfo{author}{Kamehkhosh, I.}, \& \bibinfo{author}{Jugovac, M.}
  (\bibinfo{year}{2015}).
\newblock \bibinfo{title}{What recommenders recommend: An analysis of
  recommendation biases and possible countermeasures}.
\newblock {\it \bibinfo{journal}{User Modeling and User-Adapted
  Interaction}\/},  {\it \bibinfo{volume}{25}\/}, \bibinfo{pages}{427--491}.
\bibitem[{Jannach et~al.(2010)Jannach, Zanker, Felfernig \&
  Friedrich}]{JannachZankerEtAl2010}
\bibinfo{author}{Jannach, D.}, \bibinfo{author}{Zanker, M.},
  \bibinfo{author}{Felfernig, A.}, \& \bibinfo{author}{Friedrich, G.}
  (\bibinfo{year}{2010}).
\newblock {\it \bibinfo{title}{Recommender Systems - An Introduction}\/}.
\newblock \bibinfo{address}{Cambridge, NY}: \bibinfo{publisher}{Cambridge
  University Press}.
\bibitem[{Karpinski(2005)}]{karpinski2005}
\bibinfo{author}{Karpinski, R.} (\bibinfo{year}{2005}).
\newblock \bibinfo{title}{The next phase: Bottom-up marketing}.
\newblock {\it \bibinfo{journal}{BtoB Magazine}\/},  {\it
  \bibinfo{volume}{90}\/}, \bibinfo{pages}{38}.
\bibitem[{Kazil et~al.(2020)Kazil, Masad \& Crooks}]{python-mesa-2020}
\bibinfo{author}{Kazil, J.}, \bibinfo{author}{Masad, D.}, \&
  \bibinfo{author}{Crooks, A.} (\bibinfo{year}{2020}).
\newblock \bibinfo{title}{{Utilizing Python for agent-based modeling: The Mesa
  framework}}.
\newblock In {\it \bibinfo{booktitle}{International Conference on Social
  Computing, Behavioral-Cultural Modeling and Prediction and Behavior
  Representation in Modeling and Simulation}\/} (pp.
  \bibinfo{pages}{308--317}).
\bibitem[{Kim et~al.(2009)Kim, Ferrin \& Rao}]{kim2009trust}
\bibinfo{author}{Kim, D.~J.}, \bibinfo{author}{Ferrin, D.~L.}, \&
  \bibinfo{author}{Rao, H.~R.} (\bibinfo{year}{2009}).
\newblock \bibinfo{title}{Trust and satisfaction, two stepping stones for
  successful e-commerce relationships: A longitudinal exploration}.
\newblock {\it \bibinfo{journal}{Information Systems Research}\/},  {\it
  \bibinfo{volume}{20}\/}, \bibinfo{pages}{237--257}.
\bibitem[{Koren(2008)}]{koren2008factorization}
\bibinfo{author}{Koren, Y.} (\bibinfo{year}{2008}).
\newblock \bibinfo{title}{Factorization meets the neighborhood: A multifaceted
  collaborative filtering model}.
\newblock In {\it \bibinfo{booktitle}{Proceedings of the 14th ACM SIGKDD
  International Conference on Knowledge Discovery and Data Mining}\/} (pp.
  \bibinfo{pages}{426--434}).
\bibitem[{Leitner et~al.(2017)Leitner, Rausch \& Behrens}]{leitner2017}
\bibinfo{author}{Leitner, S.}, \bibinfo{author}{Rausch, A.}, \&
  \bibinfo{author}{Behrens, D.~A.} (\bibinfo{year}{2017}).
\newblock \bibinfo{title}{Distributed investment decisions and forecasting
  errors: An analysis based on a multi-agent simulation model}.
\newblock {\it \bibinfo{journal}{European Journal of Operational Research}\/},
  {\it \bibinfo{volume}{258}\/}, \bibinfo{pages}{279--294}.
\bibitem[{Leitner \& Wall(2014)}]{leitner2014}
\bibinfo{author}{Leitner, S.}, \& \bibinfo{author}{Wall, F.}
  (\bibinfo{year}{2014}).
\newblock \bibinfo{title}{Multiobjective decision making policies and
  coordination mechanisms in hierarchical organizations: Results of an
  agent-based simulation}.
\newblock {\it \bibinfo{journal}{The Scientific World Journal}\/},  {\it
  \bibinfo{volume}{Article 875146}\/}.
\bibitem[{Leitner \& Wall(2015)}]{Leitner2015}
\bibinfo{author}{Leitner, S.}, \& \bibinfo{author}{Wall, F.}
  (\bibinfo{year}{2015}).
\newblock \bibinfo{title}{Simulation-based research in management accounting
  and control: An illustrative overview}.
\newblock {\it \bibinfo{journal}{Journal of Management Control}\/},  {\it
  \bibinfo{volume}{26}\/}, \bibinfo{pages}{105--129}.
\bibitem[{Linden et~al.(2003)Linden, Smith \& York}]{Linden2003}
\bibinfo{author}{Linden, G.}, \bibinfo{author}{Smith, B.}, \&
  \bibinfo{author}{York, J.} (\bibinfo{year}{2003}).
\newblock \bibinfo{title}{Amazon. com recommendations: Item-to-item
  collaborative filtering}.
\newblock {\it \bibinfo{journal}{IEEE Internet Computing}\/},  {\it
  \bibinfo{volume}{7}\/}, \bibinfo{pages}{76--80}.
\bibitem[{Liu et~al.(2014)Liu, Datta \& Lim}]{liu2014computational}
\bibinfo{author}{Liu, X.}, \bibinfo{author}{Datta, A.}, \&
  \bibinfo{author}{Lim, E.-P.} (\bibinfo{year}{2014}).
\newblock {\it \bibinfo{title}{Computational trust models and machine
  learning}\/}.
\newblock \bibinfo{publisher}{CRC Press}.
\bibitem[{Mladenov et~al.(2020)Mladenov, Hsu, Jain, Ie, Colby, Mayoraz, Pham,
  Tran, Vendrov \& Boutilier}]{mladenov2021recsimng}
\bibinfo{author}{Mladenov, M.}, \bibinfo{author}{Hsu, C.-w.},
  \bibinfo{author}{Jain, V.}, \bibinfo{author}{Ie, E.}, \bibinfo{author}{Colby,
  C.}, \bibinfo{author}{Mayoraz, N.}, \bibinfo{author}{Pham, H.},
  \bibinfo{author}{Tran, D.}, \bibinfo{author}{Vendrov, I.}, \&
  \bibinfo{author}{Boutilier, C.} (\bibinfo{year}{2020}).
\newblock \bibinfo{title}{{Demonstrating principled uncertainty modeling for
  recommender ecosystems with RecSim NG}}.
\newblock In {\it \bibinfo{booktitle}{Fourteenth ACM Conference on Recommender
  Systems}\/} (p. \bibinfo{pages}{591–593}).
\bibitem[{Nadolski et~al.(2009)Nadolski, Van~den Berg, Berlanga, Drachsler,
  Hummel, Koper \& Sloep}]{Nadolski2009}
\bibinfo{author}{Nadolski, R.}, \bibinfo{author}{Van~den Berg, B.},
  \bibinfo{author}{Berlanga, A.}, \bibinfo{author}{Drachsler, H.},
  \bibinfo{author}{Hummel, H.}, \bibinfo{author}{Koper, R.}, \&
  \bibinfo{author}{Sloep, P.} (\bibinfo{year}{2009}).
\newblock \bibinfo{title}{Simulating light-weight personalised recommender
  systems in learning networks: A case for pedagogy-oriented and rating-based
  hybrid recommendation strategies}.
\newblock {\it \bibinfo{journal}{Journal of Artificial Societies and Social
  Simulation}\/},  {\it \bibinfo{volume}{12}\/}, \bibinfo{pages}{1--4}.
\bibitem[{Nelson(1970)}]{nelson1970}
\bibinfo{author}{Nelson, P.} (\bibinfo{year}{1970}).
\newblock \bibinfo{title}{Information and consumer behavior}.
\newblock {\it \bibinfo{journal}{Journal of Political Economy}\/},  {\it
  \bibinfo{volume}{78}\/}, \bibinfo{pages}{311--329}.
\bibitem[{Oliver(1980)}]{oliver1980cognitive}
\bibinfo{author}{Oliver, R.~L.} (\bibinfo{year}{1980}).
\newblock \bibinfo{title}{A cognitive model of the antecedents and consequences
  of satisfaction decisions}.
\newblock {\it \bibinfo{journal}{Journal of Marketing Research}\/},  {\it
  \bibinfo{volume}{17}\/}, \bibinfo{pages}{460--469}.
\bibitem[{Pathak et~al.(2010)Pathak, Garfinkel, Gopal, Venkatesan \&
  Yin}]{pathak2010}
\bibinfo{author}{Pathak, B.}, \bibinfo{author}{Garfinkel, R.},
  \bibinfo{author}{Gopal, R.~D.}, \bibinfo{author}{Venkatesan, R.}, \&
  \bibinfo{author}{Yin, F.} (\bibinfo{year}{2010}).
\newblock \bibinfo{title}{Empirical analysis of the impact of recommender
  systems on sales}.
\newblock {\it \bibinfo{journal}{Journal of Management Information Systems}\/},
   {\it \bibinfo{volume}{27}\/}, \bibinfo{pages}{159--188}.
\bibitem[{Prawesh \& Padmanabhan(2014)}]{prawesh2014most}
\bibinfo{author}{Prawesh, S.}, \& \bibinfo{author}{Padmanabhan, B.}
  (\bibinfo{year}{2014}).
\newblock \bibinfo{title}{The ``most popular news'' recommender: Count
  amplification and manipulation resistance}.
\newblock {\it \bibinfo{journal}{Information Systems Research}\/},  {\it
  \bibinfo{volume}{25}\/}, \bibinfo{pages}{569--589}.
\bibitem[{Robertson(2016)}]{robertson2016}
\bibinfo{author}{Robertson, D.~A.} (\bibinfo{year}{2016}).
\newblock \bibinfo{title}{Agent-based models and behavioral operational
  research}.
\newblock In \bibinfo{editor}{M.~Kunc}, \bibinfo{editor}{J.~Malpass}, \&
  \bibinfo{editor}{L.~White} (Eds.), {\it \bibinfo{booktitle}{Behavioral
  Operational Research}\/} (pp. \bibinfo{pages}{137--159}).
\bibitem[{Rohde et~al.(2018)Rohde, Bonner, Dunlop, Vasile \&
  Karatzoglou}]{rohde2018recogym}
\bibinfo{author}{Rohde, D.}, \bibinfo{author}{Bonner, S.},
  \bibinfo{author}{Dunlop, T.}, \bibinfo{author}{Vasile, F.}, \&
  \bibinfo{author}{Karatzoglou, A.} (\bibinfo{year}{2018}).
\newblock \bibinfo{title}{{RecoGym: A reinforcement learning environment for
  the problem of product recommendation in online advertising}}.
\newblock \bibinfo{howpublished}{arXiv preprint arXiv:1808.00720}.
\newblock \URLprefix \url{https://arxiv.org/abs/1808.00720}.
  \href{http://arxiv.org/abs/1808.00720}{\tt arXiv:1808.00720}.
\bibitem[{Shi et~al.(2019)Shi, Ozsoy, Hurley, Smyth, Tragos, Geraci \&
  Lawlor}]{shi2019pyrecgym}
\bibinfo{author}{Shi, B.}, \bibinfo{author}{Ozsoy, M.~G.},
  \bibinfo{author}{Hurley, N.}, \bibinfo{author}{Smyth, B.},
  \bibinfo{author}{Tragos, E.~Z.}, \bibinfo{author}{Geraci, J.}, \&
  \bibinfo{author}{Lawlor, A.} (\bibinfo{year}{2019}).
\newblock \bibinfo{title}{{PyRecGym: A reinforcement learning gym for
  recommender systems}}.
\newblock In {\it \bibinfo{booktitle}{Proceedings of the 13th ACM Conference on
  Recommender Systems}\/} (pp. \bibinfo{pages}{491--495}).
\bibitem[{Sie et~al.(2010)Sie, Bitter{-}Rijpkema \& Sloep}]{SIE20102883}
\bibinfo{author}{Sie, L.~L., Rory}, \bibinfo{author}{Bitter{-}Rijpkema, M.}, \&
  \bibinfo{author}{Sloep, P.~B.} (\bibinfo{year}{2010}).
\newblock \bibinfo{title}{A simulation for content-based and utility-based
  recommendation of candidate coalitions in virtual creativity teams}.
\newblock {\it \bibinfo{journal}{Procedia Computer Science}\/},  {\it
  \bibinfo{volume}{1}\/}, \bibinfo{pages}{2883--2888}.
\bibitem[{Singh et~al.(2020)Singh, Pramanik \&
  Choudhury}]{singh2020collaborative}
\bibinfo{author}{Singh, P.~K.}, \bibinfo{author}{Pramanik, P. K.~D.}, \&
  \bibinfo{author}{Choudhury, P.} (\bibinfo{year}{2020}).
\newblock \bibinfo{title}{Collaborative filtering in recommender systems:
  Technicalities, challenges, applications, and research trends}.
\newblock In \bibinfo{editor}{G.~Shrivastava}, \bibinfo{editor}{S.-L. Peng},
  \bibinfo{editor}{H.~Bansal}, \bibinfo{editor}{K.~Sharma}, \&
  \bibinfo{editor}{M.~Sharma} (Eds.), {\it \bibinfo{booktitle}{New Age
  Analytics}\/} (pp. \bibinfo{pages}{183--215}).
\newblock \bibinfo{publisher}{Apple Academic Press}.
\bibitem[{Steinbacher et~al.(2021)Steinbacher, Raddant, Karimi, Camacho~Cuena,
  Alfarano, Iori \& Lux}]{steinbacher2021advances}
\bibinfo{author}{Steinbacher, M.}, \bibinfo{author}{Raddant, M.},
  \bibinfo{author}{Karimi, F.}, \bibinfo{author}{Camacho~Cuena, E.},
  \bibinfo{author}{Alfarano, S.}, \bibinfo{author}{Iori, G.}, \&
  \bibinfo{author}{Lux, T.} (\bibinfo{year}{2021}).
\newblock \bibinfo{title}{Advances in the agent-based modeling of economic and
  social behavior}.
\newblock {\it \bibinfo{journal}{SN Business \& Economics}\/},  {\it
  \bibinfo{volume}{1}\/}, \bibinfo{pages}{1--24}.
\bibitem[{Thies et~al.(2019)Thies, Kieckh{\"a}fer, Spengler \&
  Sodhi}]{thies2019operations}
\bibinfo{author}{Thies, C.}, \bibinfo{author}{Kieckh{\"a}fer, K.},
  \bibinfo{author}{Spengler, T.~S.}, \& \bibinfo{author}{Sodhi, M.~S.}
  (\bibinfo{year}{2019}).
\newblock \bibinfo{title}{Operations research for sustainability assessment of
  products: A review}.
\newblock {\it \bibinfo{journal}{European Journal of Operational Research}\/},
  {\it \bibinfo{volume}{274}\/}, \bibinfo{pages}{1--21}.
\bibitem[{Umeda et~al.(2014)Umeda, Ichikawa, Koyama \&
  Deguchi}]{Umeda2014EvaluationOC}
\bibinfo{author}{Umeda, T.}, \bibinfo{author}{Ichikawa, M.},
  \bibinfo{author}{Koyama, Y.}, \& \bibinfo{author}{Deguchi, H.}
  (\bibinfo{year}{2014}).
\newblock \bibinfo{title}{Evaluation of collaborative filtering by agent-based
  simulation considering market environment}.
\newblock In {\it \bibinfo{booktitle}{Developments in Business Simulation and
  Experiential Learning: Proceedings of the Annual ABSEL Conference}\/} (pp.
  \bibinfo{pages}{214--222}).
\bibitem[{Wall \& Leitner(2021)}]{Wall2020}
\bibinfo{author}{Wall, F.}, \& \bibinfo{author}{Leitner, S.}
  (\bibinfo{year}{2021}).
\newblock \bibinfo{title}{Agent-based computational economics in management
  accounting research: Opportunities and difficulties}.
\newblock {\it \bibinfo{journal}{Journal of Management Accounting Research}\/},
   {\it \bibinfo{volume}{33}\/}, \bibinfo{pages}{189--212}.
\bibitem[{Yoo et~al.(2013)Yoo, Gretzel \& Zanker}]{PersuasiveRS2013}
\bibinfo{author}{Yoo, K.-H.}, \bibinfo{author}{Gretzel, U.}, \&
  \bibinfo{author}{Zanker, M.} (\bibinfo{year}{2013}).
\newblock {\it \bibinfo{title}{Persuasive recommender systems: Conceptual
  background and implications}\/}.
\newblock \bibinfo{address}{New York, NY}: \bibinfo{publisher}{Springer}.
\bibitem[{Zajac \& Huber(2021)}]{zajac2021objectives}
\bibinfo{author}{Zajac, S.}, \& \bibinfo{author}{Huber, S.}
  (\bibinfo{year}{2021}).
\newblock \bibinfo{title}{Objectives and methods in multi-objective routing
  problems: A survey and classification scheme}.
\newblock {\it \bibinfo{journal}{European Journal of Operational Research}\/},
  {\it \bibinfo{volume}{290}\/}, \bibinfo{pages}{1--25}.
\bibitem[{Zhang et~al.(2020)Zhang, Adomavicius, Gupta \&
  Ketter}]{consumptionPerformance2020}
\bibinfo{author}{Zhang, J.}, \bibinfo{author}{Adomavicius, G.},
  \bibinfo{author}{Gupta, A.}, \& \bibinfo{author}{Ketter, W.}
  (\bibinfo{year}{2020}).
\newblock \bibinfo{title}{Consumption and performance: Understanding
  longitudinal dynamics of recommender systems via an agent-based simulation
  framework}.
\newblock {\it \bibinfo{journal}{Information Systems Research}\/},  {\it
  \bibinfo{volume}{31}\/}, \bibinfo{pages}{76–101}.
\bibitem[{Zhou et~al.(2021)Zhou, Zhang \& Adomavicius}]{longitudinalimpact2021}
\bibinfo{author}{Zhou, M.}, \bibinfo{author}{Zhang, J.}, \&
  \bibinfo{author}{Adomavicius, G.} (\bibinfo{year}{2021}).
\newblock \bibinfo{title}{Longitudinal impact of preference biases on
  recommender systems' performance}.
\newblock \bibinfo{howpublished}{SSRN Electronic Journal}.
\newblock \URLprefix \url{https://ssrn.com/abstract=3799525}.

\end{thebibliography}

\newpage
\setcounter{page}{1}
\appendix
\setcounter{figure}{0}
\section{Varying the Customer Expectation Threshold}
\label{appendix:results-varying-expectation-threshold}
Here, we report and briefly discuss the outcomes of the simulations when varying the customer expectation threshold $\psi \in \{0.75, 0.85, 0.95\}$. Social media is not considered in these simulations, i.e., $\Delta=0$. All other parameters are fixed to the values included in  Tab.~\ref{tab:model-parameters} and Tab.~\ref{tab:static-model-parameters}. Recall, in these scenarios, the consumption probability is equal to trust.

\begin{figure}[h!t]
     \centering
     \begin{subfigure}{0.3\textwidth}
         \centering
         \includegraphics [width=\textwidth]{results/social_reliance0/consumption}
         \caption{$\psi = 0.75$}
         \label{fig:with-consumer-expectation0.75-trust}
     \end{subfigure}
     \hspace{0.1em}
     \begin{subfigure}{0.3\textwidth}
         \centering
         \includegraphics[width=\textwidth]{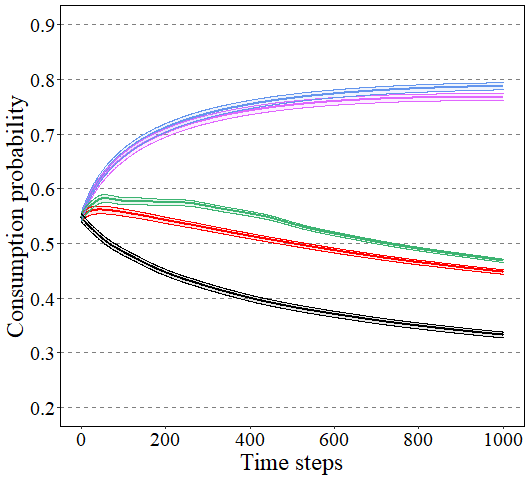}
         \caption{$\psi = 0.85$}
         \label{fig:with-consumer-expectation0.85-trust}
     \end{subfigure}
     \hspace{0.1em}
     \begin{subfigure}{0.3\textwidth}
         \centering
         \includegraphics[width=\textwidth]{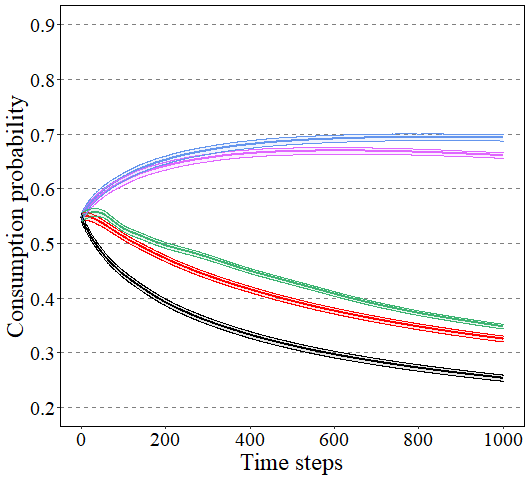}
         \caption{$\psi = 0.95$}
         \label{fig:with-consumer-expectation0.95-trust}
     \end{subfigure}
     \par\bigskip
      \begin{subfigure}{0.8\textwidth}
         \centering
         \includegraphics[width=0.8\textwidth]{results/legend}
     \end{subfigure}
        \caption{Effects of varying consumers expectations $\psi$ on their consumptions.}
        \label{fig:trust-with-consumer-expectation}
\end{figure}

Figure~\ref{fig:trust-with-consumer-expectation} shows the outcomes for the three threshold levels. We observe that the overall patterns, e.g., consumer-oriented strategies leading to the highest trust levels, remain the same, but the absolute values are generally lower. Also, higher consumer expectations lead to a more rapid decrease in trust.

\begin{figure}[h!t]
     \centering
     \begin{subfigure}{0.3\textwidth}
         \centering
         \includegraphics [width=\textwidth]{results/social_reliance0/profit-per-step}
         \caption{$\psi = 0.75$}
         \label{fig:with-consumer-expectation0.75-profit-per-step}
     \end{subfigure}
     \hspace{0.1em}
     \begin{subfigure}{0.3\textwidth}
         \centering
         \includegraphics[width=\textwidth]{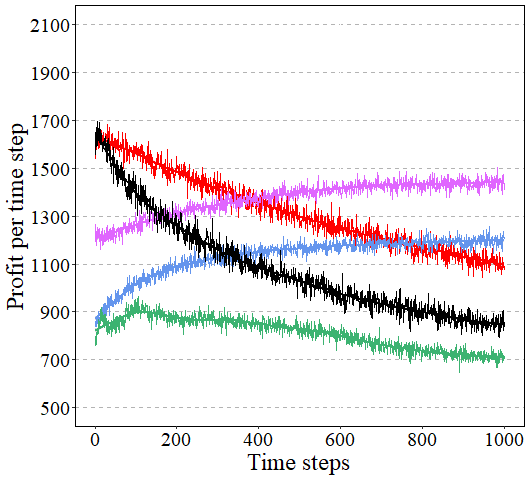}
         \caption{$\psi = 0.85$}
         \label{fig:with-consumer-expectation0.85-profit-per-step}
     \end{subfigure}
     \hspace{0.1em}
     \begin{subfigure}{0.3\textwidth}
         \centering
         \includegraphics[width=\textwidth]{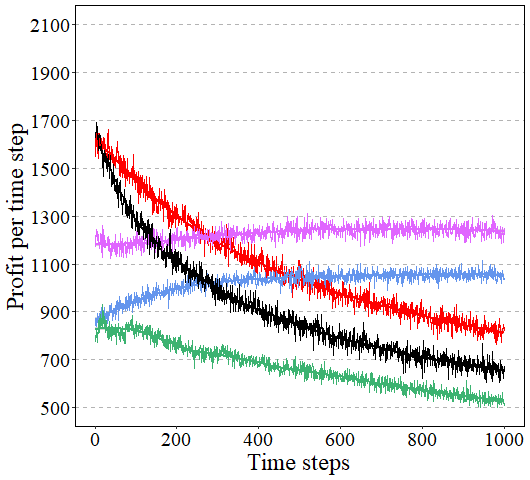}
         \caption{$\psi = 0.95$}
         \label{fig:with-consumer-expectation0.95-profit-per-step}
     \end{subfigure}
     \par\bigskip
      \begin{subfigure}{0.8\textwidth}
         \centering
         \includegraphics[width=0.8\textwidth]{results/legend}
     \end{subfigure}
        \caption{Effects of varying consumers expectations $\psi$ on the profit per time step.}
        \label{fig:profit-per-step-with-consumer-expectation}
\end{figure}

\begin{figure}[h!t]
     \centering
     \begin{subfigure}{0.3\textwidth}
         \centering
         \includegraphics [width=\textwidth]{results/social_reliance0/cumulative-profit}
         \caption{$\psi = 0.75$}
         \label{fig:with-consumer-expectation0.75-cumulative-profit}
     \end{subfigure}
     \hspace{0.1em}
     \begin{subfigure}{0.3\textwidth}
         \centering
         \includegraphics[width=\textwidth]{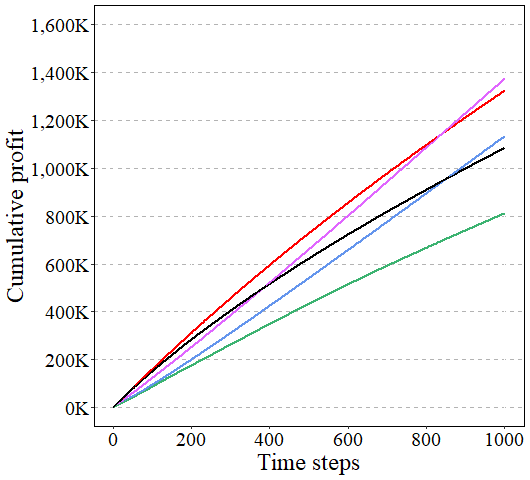}
         \caption{$\psi = 0.85$}
         \label{fig:with-consumer-expectation0.85-cumulative-profit}
     \end{subfigure}
     \hspace{0.1em}
     \begin{subfigure}{0.3\textwidth}
         \centering
         \includegraphics[width=\textwidth]{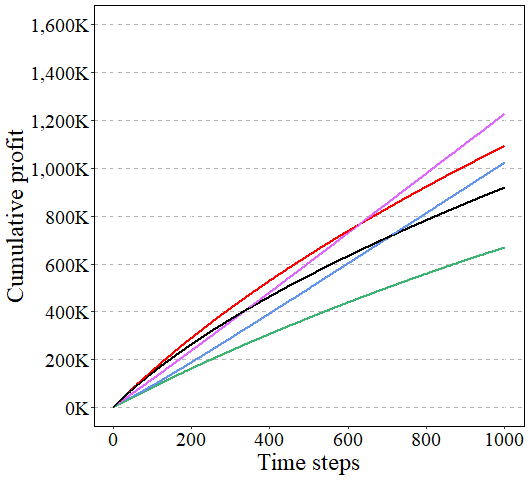}
         \caption{$\psi = 0.95$}
         \label{fig:with-consumer-expectation0.95-cumulative-profit}
     \end{subfigure}
     \par\bigskip
      \begin{subfigure}{0.8\textwidth}
         \centering
         \includegraphics[width=0.8\textwidth]{results/legend}
     \end{subfigure}
        \caption{Effects of varying consumers expectations $\psi$ on the cumulative profit.}
        \label{fig:cumulative-with-consumer-expectation}
\end{figure}

Figure~\ref{fig:profit-per-step-with-consumer-expectation} shows the average profit when varying the expectation threshold; Fig.~\ref{fig:cumulative-with-consumer-expectation} depicts the cumulative profit.
Considering the profit per time step (Fig.~\ref{fig:profit-per-step-with-consumer-expectation}), we observe trends similar to the ones for consumer trust, i.e., the absolute values are lower (due to lower consumption probabilities), and the initial drop in numbers is faster for those strategies that do not focus on consumer value. When setting the consumer expectations very high ($\psi=0.95$), the stronger decrease of the balanced strategy is very pronounced.
Looking at the cumulative profit (Figure~\ref{fig:cumulative-with-consumer-expectation}), the more rapid reduction of trust ultimately leads to the effect that the cumulative profit of the consumer-biased strategy exceeds the profit obtained with the balanced strategy at the end of the simulation. 
Thus, for recommender systems in practice, the results suggest that, in the long run, \textit{(a)} focusing on profit alone is not a feasible strategy, \textit{(b)} the higher the consumers' expectations are, the more advantageous is a consumer-biased over the other strategies, and \textit{(c)} the higher consumer expectations are, the more detrimental are non-personalized strategies to profit.

\end{document}